%% file: main_reorg.tex
\documentclass[trackchanges,twocolumn]{aastex701} 
\pdfoutput=1 
\usepackage[T1]{fontenc}
\usepackage{amsmath,amstext}
\usepackage{amssymb}
\usepackage{soul}
\usepackage[T1]{tipa} 
\usepackage{apjfonts} 
\usepackage[figure,figure*]{hypcap}
\usepackage{microtype}
\usepackage{tablefootnote}
\usepackage{booktabs}  
\usepackage{longtable}
\usepackage{hyperref}
\usepackage{CJKutf8}
\usepackage{float}
\usepackage{caption}

\def\ra#1#2#3{#1$^{\rm h}$#2$^{\rm m}$#3$^{\rm s}$}
\def\dec#1#2#3{$#1^\circ#2'#3''$}

\newcommand{\kkoname}{k'ni\textipa{P}atn k'l$\left._\mathrm{\smile}\right.$stk'masqt}

\newcommand{\ndwarfs}{five}

\begin{document}
\begin{CJK*}{UTF8}{gbsn}
\shortauthors{Dong et al.}

\title{Probing the Metallicity Dependence of Fast Radio Burst Progenitors with CHIME/FRB Outrigger Dwarf Host Galaxies}
 
\input{aff}
\input{authors}

\begin{abstract}



Using the fast radio bursts (FRBs) localized with the CHIME/FRB Outriggers between February and October 2025, we conduct a systematic search for dwarf host galaxies. We identify and characterize eight host galaxies for one repeating and seven apparently non-repeating FRBs, newly classifying \ndwarfs~hosts as dwarfs with stellar masses of $\log(M_\ast/M_\odot) \approx 8.1-9.1$. 
We measure their gas-phase metallicities and, when combined with hosts from the literature, finding that repeating and non-repeating FRBs have statistically distinct metallicity distributions. Non-repeating FRBs inhabit relatively metal-rich environments ($12 + \rm log( O/H) \gtrsim 8.35$) across the full observed mass range, while repeating FRBs have significant preference for dwarf galaxies with $12 + \rm log( O/H) \lesssim 8.35$. Almost all repeaters associated with persistent radio sources (PRSs) are in dwarfs with $12 + \rm log( O/H) \lesssim 8.15$, suggesting a connection between repeaters in metal-poor dwarf environments and PRS formation. We also calculate the host dispersion measure ($\mathrm{DM}_{\rm host}$), Faraday rotation measure ($\mathrm{RM}_{\rm host}$), and scattering timescale ($\tau_{\rm host,1~GHz}$) for 54, 37, and 24 FRBs, respectively, including both our sample and FRBs in the literature. We find no significant ($>3\sigma$) correlation between host metallicity and $\mathrm{DM}_{\rm host}$, $\mathrm{RM}_{\rm host}$, or $\tau_{\rm host,1~GHz}$. The distinct host metallicities of apparently non-repeating and repeating FRBs may reflect metallicity-dependent massive-star evolution, with repeating FRBs and PRSs requiring more extreme magnetar birth spins than the predominantly non-repeating FRB population. 

\end{abstract}


\section{Introduction}
\label{sec:intro}

Found at the lowest mass end of the galaxy mass function, dwarf galaxies are the most numerous galaxies in the Universe, yet among the most difficult to detect because of their low luminosities and surface brightnesses. Their connection to cosmic explosions came into focus through long $\gamma$-ray bursts (LGRBs) and hydrogen-poor superluminous supernovae (SLSNe-I), the only known transient populations with a strong preference for highly star-forming dwarf galaxies \citep{Frutcher06, Lunnan14, Perley16, Schulze21}. The discovery of the first well-localized fast radio burst (FRB) -- a millisecond-duration pulse of bright radio emission from cosmological distances \citep{Lorimer07} -- to a similarly star-forming dwarf galaxy \citep{Chatterjee17, Tendulkar17} therefore prompted immediate comparisons between these transient classes \citep{Murase16, Metzger17, Nicholl17}.

However, as FRB host galaxies have accumulated through several experiments including the Canadian Hydrogen Intensity Mapping Experiment (CHIME/FRB; \citealt{Michilli23, Ibik24}), the Commensal Real-Time ASKAP Fast-Transients (CRAFT; \citealt{Shannon25}) survey, the Deep Synoptic Array (DSA-110; \citealt{Law24}), and the
More TRAnsients and Pulsars project (MeerTRAP; \citealt{Pastor26}), the majority of identified hosts have been found to be typical star-forming galaxies with moderate stellar masses \citep{Gordon23, Sharma24, Shannon25}, making dwarf galaxies remarkable exceptions. Moreover, the current host demographics has led to the suggestion that FRBs preferentially reside in massive star-forming galaxies \citep{Sharma24}. Nevertheless, targeted follow-up searches have uncovered a small number of FRB dwarf host galaxies. Dwarf galaxies have been associated with both repeating FRBs, (``repeaters''; \citealt{Tendulkar17, Niu22, Bhardwaj25, Dial26, Moroianu26}), for which multiple bursts have been detected from the same source, and apparently non-repeating FRBs (``non-repeaters''; \citealt{Mahony18, Bhandari23, Caleb25, Muller26}). 

The significance of FRB dwarf hosts is underscored by the presence of luminous, compact persistent radio sources (PRSs) coincident with FRB positions. All four widely-accepted PRSs reside in dwarf hosts and have been associated with repeating FRBs. These remain the \textit{only} confirmed counterparts to the extragalactic FRB population at any wavelength and have been interpreted as emission from nebulae powered by a central engine, such as a young magnetar \citep{Murase16, Margalit18, Li20}, or emission from accreting compact objects \citep{Sridhar+22}. Their discovery has motivated numerous campaigns to identify compact radio sources in dwarf galaxies and investigate their potential connections to FRBs, raising the question of whether the dwarf host environment plays a role in PRS production \citep{Eftekhari19, Mondal20, Eftekhari21, Vohl23, Dong24b}. 

One of the stellar population properties closely related to stellar mass is the metallicity of the galaxy, and this relationship can be instrumental in understanding transient progenitors. For instance, the preference of LGRBs and SLSNe-I for metal-poor dwarf galaxies is consistent with their origins from the youngest and most massive stellar progenitors \citep{Leloudas15, Nicholl15}, in contrast to transients such as Type Ia SNe and short GRBs, which arise from both young and old stellar populations \citep{Sullivan06, Nugent22}. However, both identifying dwarf host galaxies and determining their metallicities is tricky because their low luminosities require longer integration times to detect the emission lines used for metallicity diagnostics. This challenge is especially acute for non-repeaters, where a single burst provides the only opportunity to localize to sufficient precision in order to identify and subsequently characterize the host environment. 

A recent development from CHIME/FRB to overcome the localization bottleneck is through the deployment and operation of very long baseline interferometry (VLBI) Outrigger stations \citep{Lanman24, CHIMEOutriggers, hco}. These new stations provide subarcsecond localizations that facilitate systematic searches for FRB host galaxies, down to the dwarf-mass regime, for both repeaters and non-repeaters. CHIME/FRB Outriggers are well poised for this effort, having already localized a substantial sample of FRBs with arcsecond and subarcsecond localization precision (CHIME/FRB Collaboration et al., in prep.). Because non-repeaters in dwarf galaxies have been particularly challenging to accrue, the CHIME/FRB Outriggers sample also offers strong statistical power to compare the two FRB populations within a common selection function.

Here, we present the first systematic search for dwarf galaxy hosts of FRBs using the CHIME/FRB Outriggers sample. We organize the paper as follows. In Section~\ref{sec:sample}, we discuss the selection of dwarf host candidates. We describe the imaging and spectroscopic observations used to characterize dwarf host candidates in Section~\ref{sec:obs}. In Section~\ref{sec:mass}, we confirm five of the candidates as dwarf hosts, quantify their stellar mass distributions, and place constraints on the fraction of FRBs residing in dwarf galaxies. We then investigate their metallicities and explore correlations between FRB repetition, presence of PRS, burst properties and host galaxy properties in Section~\ref{sec:Z}. The implications of our results are discussed in Section~\ref{sec:discussion}, and we summarize our findings and conclusions in Section~\ref{conclusion}. Throughout the paper, we adopt the \textit{Planck} cosmological parameters for a flat $\Lambda$CDM universe, with $H_{0}$ = 67.66 km s$^{-1}$ Mpc$^{-1}$, $\Omega_m = 0.310$, and $\Omega_{\lambda} = 0.690$ \citep{Planck18}.

\input{frb_basics}
\section{Candidate Sample Selection} \label{sec:sample}
We start with a sample of 78 distinct FRBs localized by the full CHIME/FRB Outriggers array \citep{CHIMEOutriggers} over roughly the first eight months of operations, between 2025 February and October. Most of these events will be presented in a forthcoming CHIME/FRB Outriggers catalog (CHIME/FRB Collaboration et al., in prep.). The full CHIME Outrigger array consists of three stations: \kkoname~ (KKO; \citealt{Lanman24}) located in British Columbia, a second station at Green Bank Observatory (GBO) in West Virginia, and a third at Hat Creek Radio Observatory in California (HCO; \citealt{hco}). The long baselines between CHIME and the three outriggers significantly improve FRB localizations from the $O({1'})$ scales achieved with the CHIME baseband data \citep{Michilli_2021_ApJ, Michilli23} to subarcsecond scales along the shortest axis, depending on which baselines are included in the VLBI localization. Details of the VLBI correlation/localization, radio frequency interference mitigation, and astrometric calibration procedure are provided by~\citealt{Leuong21,leung2024vlbisoftwarecorrelatorfast,Andrew25, andrew_spatial_2026}.

The FRB localization ellipses are parametrized by an R.A., Dec. centroid, a semi-major axis ($a_{\rm err}$), a semi-minor axis ($b_{\rm err}$), and an orientation angle ($\theta$) measured east of north. To ensure robust associations to host galaxies, we only consider FRBs with $a_{\rm err} < 2''$. Within this sample, the median 1$\sigma$ ellipse area is $\pi (a_{\rm err}b_{\rm err}) = 0.41~\mathrm {arcsec}^2$.

Following the same procedure as the first CHIME/FRB Outriggers host catalog \citep{KKO25}, we next associate each FRB with its most probable host galaxy using the Probabilistic Association of Transients to Their Hosts (PATH; \citealt{PATH}) framework. PATH is a Bayesian framework that evaluates the posterior probability of a host association among imaged galaxies, $P(O|x)$, by combining observables such as sky positions, angular sizes, and galaxy brightnesses. We adopt the default priors and star-galaxy classifications from \cite{KKO25} and preferentially use deep (median 5$\sigma$ depth of $m_r = 23.5$ mag) public archival imaging from the Dark Energy Camera Legacy Survey Releases 8 and 10 (DECaLS DR8, 10; \citealt{Decals}). For FRBs outside the DECaLS footprint, we rely on the shallower (median 5$\sigma$ depth of $m_r = 23.2$ mag) Pan-STARRS Data Release 2 (PS1 DR2; \citealt{Chamber16, PS1}), which completely covers the CHIME/FRB sky. 

We identify the galaxy candidate with the highest $P(O|x)$. For this work, we consider the host association robust when $P(O|x)\geq0.9$. For eight FRBs that are initially ``host-less'' with an unseen host probability $P(U|x) > 60\%$ in archival imaging, and 10 FRBs whose primary host candidate has a reported $r$-band magnitude fainter than the nominal survey depth, we initiate deep, ground-based optical imaging (see Section~\ref{subsec:deepimaging}) and subsequently re-run PATH with these new observations. Since the unseen-host prior, $P(U)$, depends on the imaging depth, and our typical $3\sigma$ depth varies (but is usually deeper than $r \approx$ 25 mag), we instead calculate a custom $P(U)$ following \cite{James2026}. 


To calculate the custom $P(U)$, we use
\begin{equation}  \label{PU}
P(U \mid m \ge m_{\min}) =
\frac{\int_{m_{\min}}^{\infty} P(U \mid m)\, P(m)\, dm}
{\int_{m_{\min}}^{\infty} P(m)\, dm}
\end{equation} 
where $m_{\min}$ is 23 and 21 for DECaLs and Pan-STARRS, respectively. Here, $P(U|m)\,dm$ is the fraction of galaxies that are unseen in the catalog as a function of apparent magnitude $m$, and $P(m)$ is the magnitude distribution for CHIME/FRB, the latter of which we take from Andersen et al. (in prep.). We estimate $P(U\mid m)$ from galaxy number counts (per square degree per magnitude) measured in several images with the same total exposure time and under similar observing conditions. We differentiate the cumulative galaxy counts with respect to magnitude to obtain the observed differential counts, $N_{\rm seen}(m)$, and compare these with $N_{\text{host}}(m)$, the empirical galaxy number counts \citep{driver+16}. The ratio $N_{\text{seen}} / N_{\text{host}}$ therefore gives the fraction of galaxies detected at a given magnitude, such that $P(U\mid m) = 1 - N_{\text{seen}} / N_{\text{host}}$. For $r$-band, we calculate a custom $P(U)=0.09$ for DECaLS and $P(U)=0.04$ for Pan-STARRS, depending on the survey the initial image was obtained. Of the 18 FRBs with deep imaging, nine resulted in a $P(O|x)\geq0.9$ through custom PATH analysis.

In most cases, the stellar masses of the host galaxies are not available in our parent sample of 78 CHIME/FRB Outrigger FRBs. To select dwarf host candidates, we therefore use the galaxy $r$-band luminosity to make an initial selection on dwarf hosts, and impose a permissive threshold of $\log(L/L_{\odot}) < 9.5$. This threshold is chosen to accommodate the range of mass-to-light ($M/L$) ratios, $\sim$0.5--3, expected to vary with the galaxy stellar mass \citep{Bell03, Du20}, minimizing the risk of excluding \textit{bona fide} dwarfs at the expense of retaining some more massive galaxy interlopers. Our luminosity cutoff would only become incomplete for dwarf galaxies with exceptionally low $M/L$ ratio ($\le0.3$). To calculate luminosities, we restrict the sample to hosts with a spectroscopic redshift obtained through our follow-up programs. Our final sample consists of eight dwarf host candidates satisfying the criteria (i) a host association of $P(O|x)$ $\ge$ 0.9, (ii) a spectroscopic redshift, and (iii) an $r$-band galaxy luminosity of $\log(L/L_{\odot}) < 9.5$. The basic FRB burst properties and host information for the eight dwarf candidates are presented in Table~\ref{tab:basics}. We note that an updated PATH framework \citep{James2026} became available during the completion of this work, which incorporates information on the probability of redshift given the FRB dispersion measure (DM). However, since most of our sources are nearby and have relatively low DMs (Table~\ref{tab:frb_radio}), we do not expect this additional information to significantly alter the PATH results.

\begin{figure*}[!t]
\centering
\includegraphics[width=0.58\textwidth]{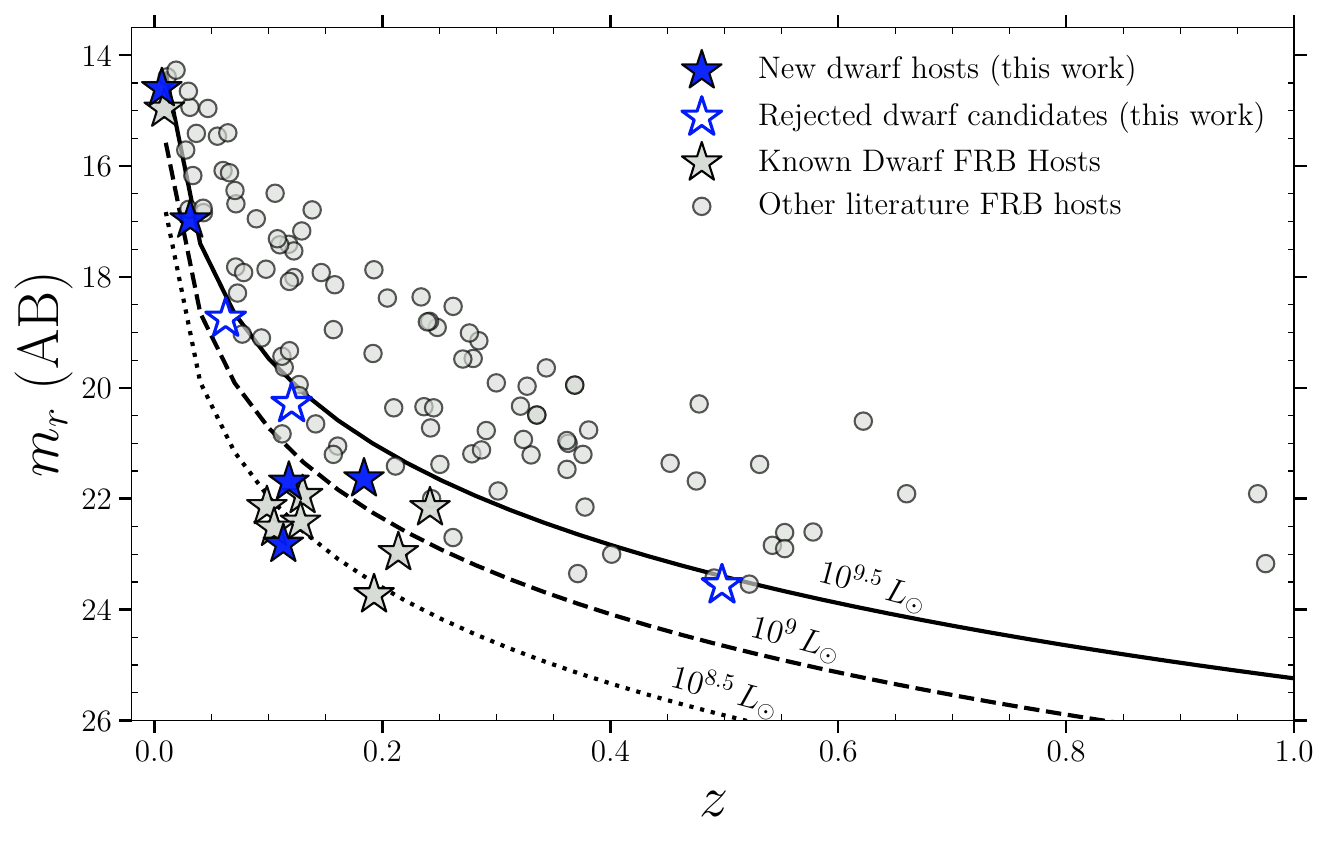}
\includegraphics[width=0.38\textwidth]{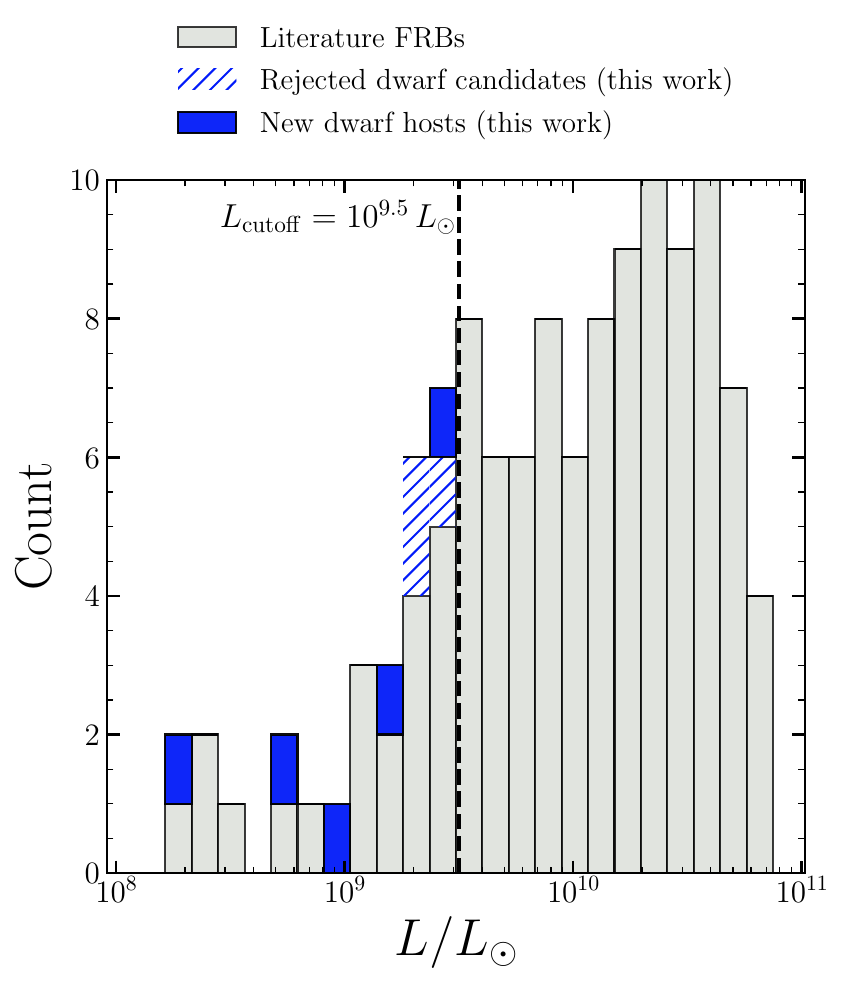}
\caption{{\it Left:} Apparent $r$-band magnitude versus redshift of FRB host galaxies overlaid with lines of constant luminosity. The confirmed dwarf galaxies in the CHIME/FRB Outriggers sample are shown as blue stars and the rejected dwarf candidates are shown as open blue stars. Previously known dwarf galaxies and the broader FRB hosts are shown as grey stars and circles, respectively (see Section~\ref{sec:sample} for references). {\it Right:} Distribution of $r$-band luminosities of all literature FRB hosts with redshifts. The vertical dashed line marks the luminosity cutoff of $\log(L/L_{\odot}) < 9.5$ used to select our dwarf galaxy candidates. The solid blue bins represent the \ndwarfs~new dwarf hosts $\log(M_{\ast}/M_{\odot}) \leq 9.1$ in this sample and the hatched bins correspond to the rejected dwarf candidates with stellar masses $\log(M_{\ast}/M_{\odot}) > 9.1$. 
\label{fig:brightness}}
\end{figure*}

In Figure~\ref{fig:brightness}, we show the $r$-band magnitude as a function of redshift, with lines of constant luminosity overlaid. We highlight the dwarf hosts and dwarf candidates rejected based on their stellar masses of $\log(M_{\ast}/M_{\odot}) > 9.1$ in this work (see Section~\ref{sec:dwarfidentification}). In this figure, we compare the hosts presented in this work, all known FRBs localized to dwarf galaxies \citep{Bhandari23, Gordon23, Lee-Waddell23, Bhardwaj25, Dial26, Moroianu26, Muller26}, and the broader FRB host population \citep{Ravi19, Marcote20, Bhandari22, Kirsten22, Bhardwaj24, Ibik24, Law24, Rajwade24, Sharma24, KKO25, Connor25, Eftekhari25, Leung25, Shannon25}.

Approximately $75\%$ of all FRB hosts lie above our luminosity cutoff of $\log(L/L_{\odot}) = 9.5$, while by design, the known dwarf hosts and candidates lie entirely below it. However, we caution that this distribution is subject to observational biases. For our parent sample of 78 FRBs, the optical host sample is incomplete, as observations are still ongoing, and some identified dwarf candidates have yet to be confirmed; see Section~\ref{subsec:dwarffrac} for details. The dwarf galaxies further separate into two regimes: bright and very nearby, with $m_r < 18$ mag and $z\lesssim0.05$, or faint and more distant with $m_r > 20$ mag and $z\gtrsim0.1$. This partition demonstrates a challenge in identifying dwarf host galaxies: while they are relatively straightforward to discover and associate with an FRB in the nearby Universe, they often require dedicated and deeper imaging to uncover at greater distances and solidify the association to an FRB. Figure~\ref{fig:brightness} also shows that luminosity alone cannot uniquely distinguish dwarf galaxies from more massive systems as the rejected dwarf candidates occupy a similar luminosity range as the confirmed dwarf hosts, though they are slightly brighter on average (see Section~\ref{sec:mass} for this further down-selection).

\section{Observations} \label{sec:obs}

\subsection{Archival Imaging and Photometry} \label{subsec:archivalimaging}

To gather multi-wavelength data to enable stellar population modeling (Section \ref{subsec:prospector}), we first search for archival photometry from optical and near-infrared (NIR) surveys. We draw from DECaLS DR10, the DECam Local Volume Exploration Survey (DELVE; \citealt{DELVE}), PS1 DR2, the Two Micron All Sky Survey AllSky (2MASS; \citealt{Twomass}), and the Wide-field Infrared Survey Explorer (WISE; \citealt{Lang16}). For the WISE data, the standard aperture radius is 8.25\arcsec~in W1, W2, W3, and 16.50\arcsec~in W4, which does not adequately capture the angular size of the host galaxies of FRB\,20250227A and 20250316A. We visually inspect the WISE images and choose the aperture that best encapsulates the host galaxy light for each source. We therefore select photometry from `aperture 5' (16.5\arcsec~in W1, W2, W3, and 33\arcsec~in W4) for FRB\,20250227A and from `aperture 8' (24.75\arcsec~in W1, W2, W3, and 49.50\arcsec~in W4) for FRB\,20250316A.

We perform manual photometry for the dwarf hosts of FRBs\,20250206A, 20250227A, and 20250316A with archival imaging where the reported photometry is unreliable, mostly due to an insufficiently small aperture. We query the archival images from the Dark Energy Spectroscopic Instrument (DESI) Legacy Imaging Survey DR 10. We reduced the archival imaging data with the \texttt{POTPyRI}
\footnote{\url{https://github.com/CIERA-Transients/POTPyRI}} pipeline. The pipeline performs standard image calibration, including bias subtraction, flat-fielding, and dark current correction (when applicable), prior to image alignment and stacking. Astrometric alignment is carried out in two stages: an initial solution using \texttt{astrometry.net}, followed by refinement through centroiding on astrometric standard stars from Gaia DR3 \citep{GaiaDR3}. Satellite trails, cosmic rays, and bad pixels are identified, masked, and excluded from the final stacked images.

Our custom script utilizes the \texttt{aperture\textunderscore photometry} function as part of the \texttt{photutils} package \citep{photutils}. We determined the appropriate aperture size and annulus by visually inspecting the host in the images using \texttt{SAOImageDS9} \citep{DS9}. For DECaLS, the photometric zero-point is fixed at 22.5~mag for all bands by definition of the nanomaggy flux unit used in the DESI Legacy Survey coadds. Finally, we correct for Galactic dust extinction for all photometric points using Galactic reddening maps from \cite{Schlafly11}. The complete photometric data are presented in Table~\ref{tab:phot}.

\subsection{Deep Ground-based Imaging and Photometry} \label{subsec:deepimaging}

Two FRBs in our sample required follow-up optical imaging observations. Specifically, FRB\,20250410E has a primary host fainter than the $r$-band  DECaLS depth, and FRB\,20250704A was initially ``host-less'' in archival imaging. We therefore initiated deep ground-based $r$-band imaging with Gemini Multi-Object Spectrographs (GMOS) on the 8-m Gemini-South telescope (GS-2025B-LP-110; PI: Eftekhari). For FRB\,20250704A, we also obtained $J$-band imaging from the MMT and Magellan infrared spectrograph (MMIRS) on the 6.5-m MMT telescope (UAO-G388-26A; PI: Eftekhari). 

For data reduction and co-addition, we utilized the \texttt{POTPyRI} pipeline, following the same procedure as for the archival imaging. The pipeline performs both aperture photometry on sources detected in the stacked image. If the host is smaller than the point spread function of the image, we use the $\texttt{Photutils}$ package to measure the flux of  stars using a range of aperture sizes. We use these measurements to create a curve of growth, which shows how much of the total flux is captured by each aperture. This allows us to determine the appropriate aperture for each host. After deriving the aperture photometry, the script determines the photometric zero-point by matching the photometry to a catalog of standard stars from supported calibration catalogs including PS1, the Sloan Digital Sky Survey (SDSS), and SkyMapper. The Galactic extinction-corrected photometry from our deep imaging observations is also presented in Table~\ref{tab:phot}.

\subsection{Spectroscopy \label{subsec:specs}}

We obtained long-slit spectroscopic observations utilized for both redshift measurement and host galaxy modeling (Section ~\ref{subsec:prospector}). For three of the dwarf hosts and candidates (FRBs\,20250227A, 20250507A, and 20250704A), we acquired Gemini/GMOS spectroscopy through our Large and Long program (GN-2025A-LP-110, GN-2025B-LP-110, GS-2025B-LP-110; PI: Eftekhari) using the B480 disperser grating and a 1\arcsec~slit. We obtained total exposure times of 2400~s for FRBs\,20250227A and 20250507A and 3600~s for FRB\,20250704A to accommodate their respective faintness. The observations were centered at 6400\AA~and 6500\AA~with a continual spectral coverage between $\approx 4400-8500$\AA. The two different central wavelengths are to cover the two small gaps due to the long-slit setup.

For FRBs\,20250206A, 20250410E, and 20250515A, we observed the host galaxies with the Low Resolution Imaging Spectrometer (LRIS) on the 10 m Keck I telescope (O397, U238; PIs: Liu, Prochaska). Both observations used a 1\arcsec~slit and the D560 dichroic. For FRB\,20250410E, we obtained six 900-s exposures using the 400/3400 grism on the blue arm and the 400/8500 grating on the red arm. For FRBs\,20250206A and 20250515A, we obtained a single 900-s exposure each using the 600/4000 grism on the blue arm and the 600/7500 grating on the red arm.

Two FRB hosts have publicly available spectroscopic observations. For FRB\,20250316A, we retrieved the galaxy spectrum from SDSS DR13 \citep{SDSS13}. For FRB\,20220529A, we used observations obtained with the Optical System for Imaging and low-Intermediate-Resolution Integrated Spectroscopy (OSIRIS+) instrument on the 10.4 m Gran Telescopio Canarias (GTC) from \citet{Li26} in which the reduced spectrum is publicly available\footnote{\url{https://github.com/Astroyx/FRB20220529}}. 

We reduced the spectroscopic data obtained from our follow-up campaigns using the Python Spectroscopic Data Reduction Pipeline (\texttt{PypeIt}; \citealt{pypeit:joss_arXiv, pypeit:zenodo}) v1.16-18. The \texttt{PypeIt} reduction procedure includes bias subtraction, flat-field correction, cosmic-ray masking, and wavelength calibration of the raw frames. The pipeline produces calibrated 2D spectral frames and extracts coadded 1D spectra from all the individual exposures. We flux-calibrated the spectra using sensitivity functions generated from standard star observations obtained on the same night whenever possible. Because standard stars for Gemini observations are typically observed only a few times per semester, we applied the standard sensitivity functions corresponding to those specific observing runs. We then applied telluric corrections to the coadded, flux-calibrated 1D spectra using a telluric absorption model spanning 3000--26,000\AA\ with a spectral resolution of $\lambda/\Delta \lambda = 15,000$. Finally, we fit the spectra with galaxy templates using \texttt{Marz} \citep{Marz} to derive redshifts, which were subsequently verified via manual inspection of the emission and absorption lines. We find a range of redshifts of $z\approx 0.06-0.5$, which are listed in Table~\ref{tab:basics}.

\input{hostproperties}

\section{Stellar Masses}  \label{sec:mass}

\subsection{SED Modeling} \label{subsec:prospector}
To derive the stellar population properties of the dwarf host galaxy sample, we modeled the galaxy spectral energy distributions (SEDs) using the Bayesian inference framework \texttt{Prospector} \citep{Johnson21}. We jointly fit the photometric and spectroscopic data to maximize the constraints on the model parameters. Many FRB host samples have been modeled with \texttt{Prospector} (e.g., \citealt{Gordon23, Sharma24}), making our results more directly comparable to the literature without introducing systematic offsets from differing modeling approaches. \texttt{Prospector} samples the posterior distribution of various stellar population properties using the dynamic nested sampling algorithm implemented in \texttt{dynesty} \citep{Speagle20}. \texttt{Prospector} builds model SEDs using the Flexible Stellar Population Synthesis library (\texttt{FSPS}; \citealt{Conroy09, Conroy10}) through the \texttt{python-fsps} interface. It also uses the \texttt{MILES} stellar library \citep{MILES11} and WMAP9 cosmology \citep{Hinshaw13} internally.


We model each galaxy star formation history (SFH) using the nonparametric \texttt{continuity} prior. From the joint posterior distribution inferred by \texttt{Prospector}, we derive several key stellar population properties: a mass-weighted age (t$_m$) which provides a more robust estimate than light-weighted age since it does not overweigh the light from young, bright stars, the recent star formation rate (SFR) averaged over the past 100~Myr (${\rm SFR}_{\rm 0-100 Myr}$), the stellar mass formed (M$_{*}$) derived from the total mass formed (M$_{F}$), the dust attenuation from young (A$_{\rm V, young}$) and old (A$_{\rm V, old}$) stars, and the stellar (Z$_*$) and gas-phase ($Z_\textrm{gas}$) metallicities. To compute oxygen abundances (Section~\ref{subsec:gasphaseZ}), we also draw 1000 representative samples of the emission-line fluxes from the posterior distribution, assuming Gaussian line profiles with a shared velocity dispersion. Additional details on the implementation of \texttt{Prospector} for the SED modeling are provided in Appendix~\ref{appendix:SEDfits}. We report the median and $68\%$ confidence intervals of the stellar population properties in Table~\ref{tab:host_prop} and display the SED fits in Figure \ref{fig:seds} for all eight candidates. 

\begin{figure*}[t!]
    \centering
    \includegraphics[width=\textwidth]{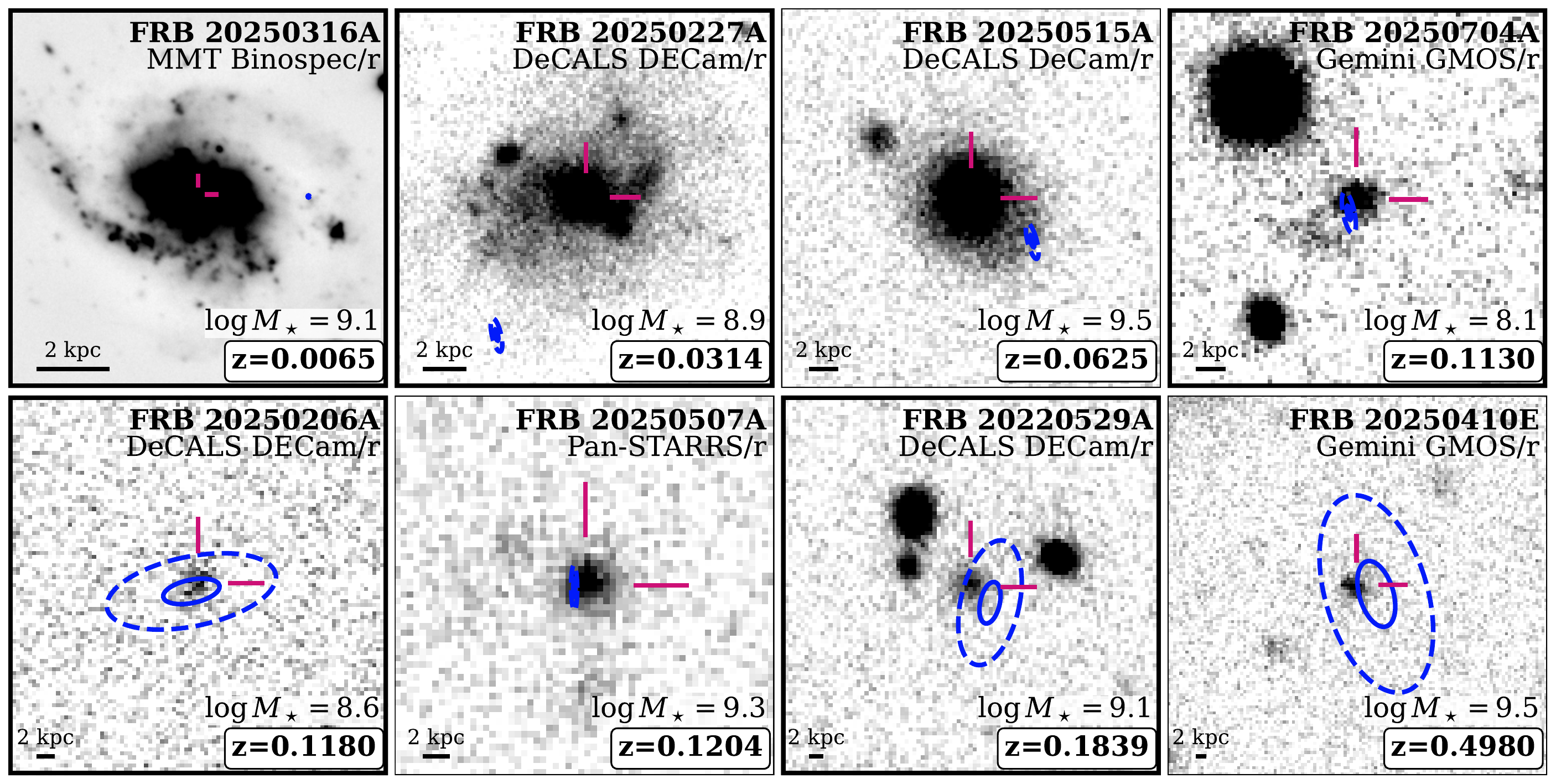}
    \caption{Image gallery of FRB dwarf host galaxies in bold outlines along with dwarf candidates rejected based on their stellar masses of $\log(M_{\ast}/M_{\odot}) > 9.1$. They are ordered by redshift, including the previously presented hosts of FRBs\,20220529A and 20250316A \citep{ RBFLOAT,Li26}, which we newly classify as dwarf galaxies. The galaxies are marked by the pink cross hairs while the 1- and 3-$\sigma$ confidence interval FRB localizations are indicated by the blue ellipses. In each panel, north is up and east is left. The survey or facility from which each optical image was obtained, along with the host redshift and derived stellar mass, are also listed.}
    \label{fig:gallery}
\end{figure*}

\subsection{Dwarf Host Identification} \label{sec:dwarfidentification}

For this work, we consider a galaxy with stellar mass of $\log(M_{\ast}/M_{\odot}) \leq 9.1$ within 1$\sigma$ uncertainty to be a dwarf. We now identify which of the eight dwarf host candidates are genuine dwarf galaxies. We determine the median and 16th/84th percentiles of the stellar masses by resampling the full posterior distribution into 100,000 uniformly-weighted samples for each candidate. Based on the resulting stellar mass distributions, we identify \ndwarfs~hosts as dwarf galaxies, with FRB\,20250316A being the most massive at $\log(M_{\ast}/M_{\odot}) = 9.1\pm 0.03$. Our derived stellar mass for FRB\,20250316A is comparable to previously reported value of $\log(M_{\ast}/M_{\odot}) = 9.2$ from the NASA/IPAC Extragalactic Database (NED) Local Volume sample \citep{Cook23}. In Figure~\ref{fig:gallery}, we present a gallery of confirmed dwarf hosts and hosts rejected as dwarfs based on their stellar masses, ordered by redshift. The \ndwarfs~confirmed dwarf hosts are: FRBs\,20220529A, 20250206A, 20250227A, 20250316A, and 20250704A; these hosts are in bold in Table~\ref{tab:basics}.

As apparent from Figure~\ref{fig:gallery}, nearby dwarf galaxies are bright enough to be readily detected and resolved, but as the distance increases, they become increasingly difficult to detect without dedicated deep follow-up observations. We find that the dwarf host sample spans log($M_*/M_\odot$) $\simeq$ 8.1--9.1 (Table~\ref{tab:host_prop}). The rejected dwarf candidates from the initial luminosity cut overall occupy a higher luminosity range, consistent with their higher inferred stellar masses between log($M_*/M_\odot$) $\simeq$ 9.3--9.5. A few known dwarf hosts are somewhat lower in mass. For instance, the host of FRB\,20190417A has $\log(M_*/M_\odot)=7.8$ \citep{Moroianu26}. Moreover, the host of FRB\,20240304B has $\log(M_*/M_\odot)=6.9$ and originates at a much higher redshift ($z\sim2$), where galaxies are expected to be lower mass overall \citep{Caleb25}. 

We further find that $t_{\rm m}$ and ${\rm SFR}_{\rm 0-100 Myr}$ of our dwarf sample span $\sim5-7$~Gyr and 0.02--0.9~$M_\odot$~yr$^{-1}$, respectively, both of which are consistent with the known dwarf host population at comparable redshifts. All known dwarf FRB hosts exhibit ${\rm SFR}_{\rm 0-100 Myr} < 0.1$ $M_\odot$~yr$^{-1}$, with the exception of the host of FRB\,20190417A at 0.19 $M_\odot$~yr$^{-1}$ inferred from H$\alpha$ fluxes \citep{Moroianu26}.

\subsection{Comparisons to Other Transients} \label{subsec:Mcomparisons}

\begin{figure*}
    \centering
    \includegraphics[width=0.85\textwidth]{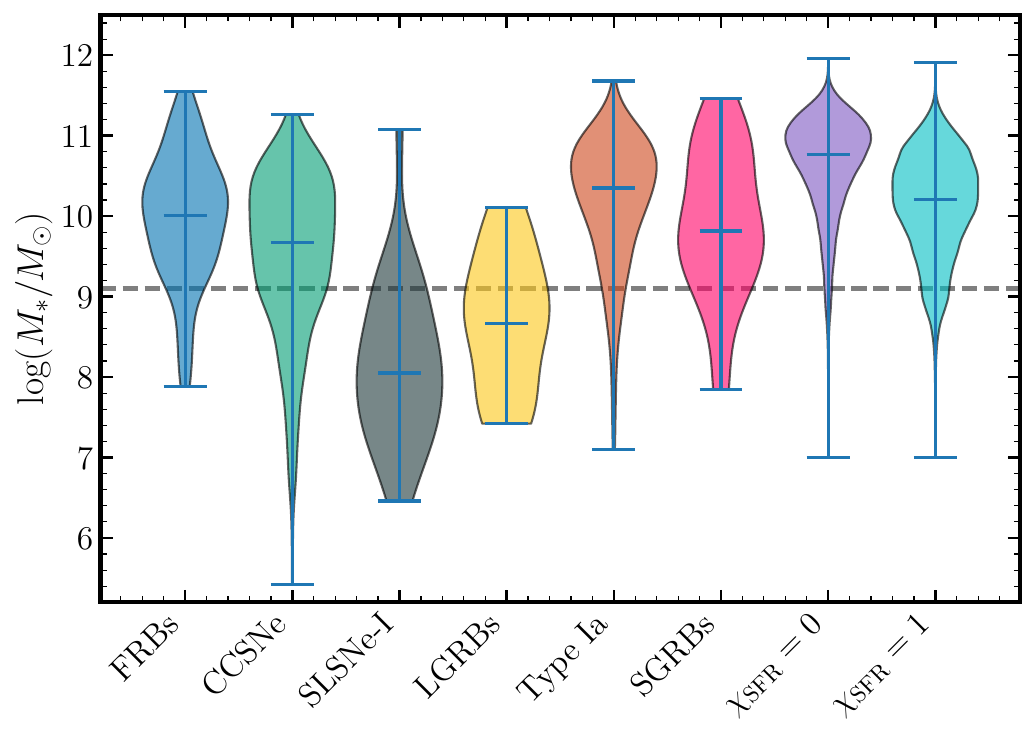}
    \caption{Stellar mass distributions of the hosts of FRBs, CCSNe, SLSNe-I, LGRBs, Type~Ia SNe, and SGRBs at $z<0.5$. We also include simulated galaxy distributions that are either mass- ($\chi_{SFR} =0$) or SF-weighted ($\chi_{SFR}=1$) generated from \texttt{GALFRB} \citep{Loudas25}. The horizontal dashed line denotes our threshold for dwarf galaxies at $\log(M_{\ast}/M_{\odot}) = 9.1$. The FRB host sample includes the new hosts presented in this work. The width of each violin is proportional to the estimated probability density of host stellar masses. The blue horizontal line represents the median stellar mass. The upper and lower caps span the full range of stellar masses (minimum to maximum) of each transient population. While these samples are not uniformly selected, it is clear that SLSNe-I and LGRBs occur in more low-mass galaxies relative to FRBs, SGRBs, CCSNe, and Type~Ia SNe. The FRB stellar mass distribution more closely resemble the SF-weighted than the M$_{*}$-weighted mock population.}
    \label{fig:stellarmass}
\end{figure*}

To contextualize the stellar mass distribution of our dwarf and rejected dwarf candidate samples within the broader FRB host population, we compare FRB hosts with those of transient populations associated with both recent star formation and older stellar populations. In Figure~\ref{fig:stellarmass}, we show violin plots of the stellar mass distributions for FRB hosts, including our sample. Our comparison includes core-collapse supernovae (CCSNe; \citealt{Schulze21}), SLSNe-I, \citep{Schulze21}, and LGRBs \citep{Taggart21}, all of which are connected to the deaths of young, massive stars. The CCSNe and SLSNe-I samples are drawn from the Palomar Transient Factory (PTF), comprising 774 CCSNe and 31 SLSNe-I discovered between 2009 and 2017 with redshifts $z<0.5$. For CCSNe, we include both stripped-envelope and hydrogen-rich events (Types Ib/c, II, IIb, and IIn). The LGRB sample consists of 17 events at $z<0.3$ discovered prior to the end of 2017. To represent progenitor channels with delayed timescales relative to star formation, we use Type Ia SNe \citep{Lampeitl10, Uddin20} and SGRBs \citep{Nugent22}. The Type Ia sample is a combination of 162 events from the SDSS-II Supernova Survey at $z<0.21$ and 113 events from the Carnegie Supernova Project-I (CSP-I) with a median redshift of $z=0.025$. The SGRB sample is restricted to $z<0.5$.

It is worth noting that all of these samples are not uniformly selected in terms of their detection criteria or host galaxy identifications. Depending on the specific FRB host catalog, galaxies are either identified through magnitude-limited surveys or dedicated, deeper observational follow-up. Much like the FRB hosts, the comparison transient host samples are constructed from various discovery surveys or instruments, meaning that their observed host properties are influenced by selection effects and detection biases associated with how each transient class is discovered and how hosts are identified. These effects preclude a direct interpretation of the observed distributions as unbiased representations of the underlying populations. However, the relative differences in the stellar mass distributions among transient classes remain evident and are difficult to attribute to selection effects alone. In Figure~\ref{fig:stellarmass}, SLSNe-I and LGRBs regularly occupy lower-mass galaxies, whereas FRBs, SGRBs, CCSNe, and Type Ia SNe span a broader range of stellar masses.

FRBs have been shown to reside primarily in galaxies with active star formation \citep{Heintz20,Bhandari22, Gordon23,Sharma24}, but there are several that instead originate from quiescent galaxies with little sign of ongoing star formation \citep{Ravi19, Gordon23, Sharma23, Shah25,Eftekhari25}, including the old stellar environment of a globular cluster \citep{Kirsten22}. This indicates that FRBs may arise from multiple progenitor systems or formation channels that are tied to both old and young stellar populations \citep{Margalit19}, often parameterized as sub-populations tracing star formation versus stellar mass in galaxies \citep{Loudas25,Horowicz26}. 

In additional to the various transients, we generate mock host galaxy samples using \texttt{GALFRB}\footnote{\url{https://github.com/loudasnick/GALFRB}} following the prescription outlined by \cite{Loudas25}. The mock galaxy population is volume-limited to $z<0.5$ to match the observed transient classes. To mimic the optical selection biases in identifying dwarf galaxies, we apply a magnitude cut of $m_r = 24$. We then construct the stellar mass distributions for populations that are purely M$_{*}$-weighted ($\chi_{\mathrm{SFR}} = 0$) or a purely SF-weighted ($\chi_{\mathrm{SFR}} = 1$). We find that the observed FRB stellar mass distribution is more consistent with the SF-weighted population with a higher fraction of low-mass galaxies, than the M$_{*}$-weighted population, for which most the of stellar mass is found at between $10 <$ $\log(M_*/M_\odot)$ $< 11$.


\subsection{Dwarf Fractions} \label{subsec:dwarffrac}

We next use this new sample of dwarf FRB hosts to place constraints on the low end of the FRB host galaxy mass function. We calculate an observed dwarf fraction defined as 
\begin{equation}
f_{\rm dwarf} = \frac{N_{\rm dwarf}}{N_{\rm total}}
\end{equation}
\noindent where $N_{\rm dwarf}$ is the number of hosts with $\log(M_{\ast}/M_{\odot}) \leq 9.1$ and $N_{\rm total}$ is the number of CHIME/FRB Outrigger FRBs at $z<0.2$, the volume where we are reliably sensitive to dwarf galaxies (i.e., Figure~\ref{fig:brightness}). Out of an $N_{\rm total}=57$ CHIME/FRB Outrigger hosts which we know to be at $z\leq 0.2$, we measure an observed fraction of $f_{\rm dwarf}>0.09$ with a 68$\%$ confidence range of $\approx0.06-0.13$ using a Wilson score interval, reflecting the modest sample size. This fraction sets a lower limit on $f_{\rm dwarf}$, since we do not yet have stellar mass measurements for all 57 CHIME/FRB Outrigger FRBs that make up the denominator of this fraction. Our derived lower limit on $f_{\rm dwarf}$ differs from that of \cite{Sharma24}, who found no dwarf FRB hosts at $z\leq 0.2$, with stellar masses below $\log(M_{\ast}/M_{\odot}) \lesssim 9.1$. This absence of dwarf FRB hosts was interpreted as a low-mass cutoff in the FRB host population, which our detection of dwarf hosts at these redshifts does not support. Moreover, there are additional candidates in the CHIME/FRB Outriggers sample that either lack spectra (and thus redshifts) or appear ``host-less'' in archival imaging aside from the dwarf candidates that we rejected in this work. As these host galaxies become characterized with future observations, this fraction may increase. If we instead use all FRB hosts from the literature at $z\leq 0.2$ including the eight presented here, we find a $f_{\rm dwarf}$ $=0.15$.

We next compare these dwarf fractions to those of known transient classes. The CCSNe, SLSNe-I, LGRB, and Type Ia samples are the same as those introduced in Section~\ref{subsec:Mcomparisons} and are similarly restricted to $z<0.2$ to provide direct comparison to our FRB population, and where selection effects on dwarf host detection should be less severe. Within this volume, we find that the observed $f_{\rm dwarf}$ for SLSNe-I (0.75) and LGRBs (0.47), are noticeably higher than both the CHIME/FRB lower limit and $f_{\rm dwarf}$ including the literature, likely owed to their specific preference for metal-poor galaxies (e.g., \citealt{Lunnan14, Perley16}). CCSNe are direct tracers of star formation and occur across both low- and high-mass galaxies, with host mass functions consistent with those of the general star-forming galaxy population weighted by star formation activity \citep{Schulze21}. As CCSNe explode in galaxies across a wide range of metallicities, the $f_{\rm dwarf}$ for CCSNe (0.3) is lower than that for SLSNe-I and LGRBs, but it is closer to the the lower limit and fraction derived for all FRB hosts. The observed $f_{\rm dwarf}$ for FRBs is the most consistent with Type Ia SNe (0.12), which trace both star formation and stellar mass.

\begin{figure}[t!]
    \centering
    \includegraphics[width=0.45\textwidth]{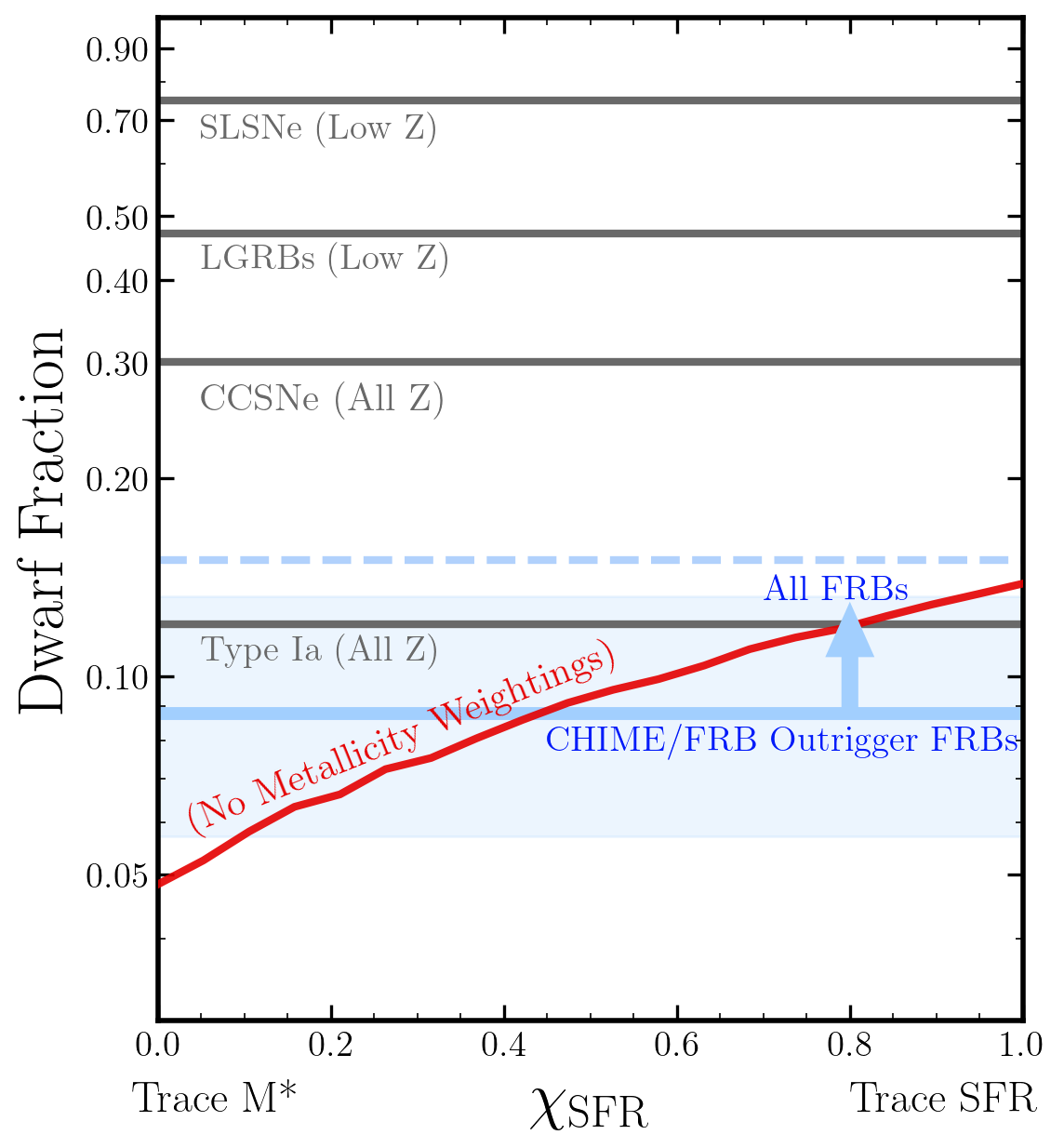}
    \caption{Lower limit on the FRB $f_{\rm dwarf}$ derived from the \ndwarfs~dwarf hosts identified in this work, together with the observed dwarf fraction of all FRB hosts, compared with other transient populations and theoretical model predictions. These fractions are restricted to a volume of $z<0.2$ where dwarf galaxies can be reliably identified. The blue shading marks the 68$\%$ confidence interval and the upward arrow indicates that this fraction is a lower limit. Gray lines correspond to the $f_{\rm dwarf}$ of SLSNe-I, LGRBs, CCSNe, and Type Ia SNe that trace either low metallicity ($Z$) or a wide range of $Z$, after applying the same stellar mass cut. The red curve shows the $f_{\rm dwarf}$ predicted for mock galaxy populations following \cite{Loudas25}, as a function of $\chi_{\mathrm{SFR}}$, which sets the relative weighting between M$_{*}$ and star formation. The derived limit on FRB $f_{\rm dwarf}$ ($0.09_{-0.03}^{+0.04}$) constrain $\chi_{\mathrm{SFR}}>0.4$, while the current FRB $f_{\rm dwarf}$ (0.15) is slightly above $\chi_{\mathrm{SFR}}=1$, suggesting FRB progenitors could purely trace recent SF, like CCSNe.}
    \label{fig:dwarffrac}
\end{figure}


We also compute $f_{\rm dwarf}$ as a function of $\chi_{\mathrm{SFR}}$, the mixing parameter that interpolates between a M$_{*}$-weighted and SF-weighted population. The mock galaxy population is volume-limited to $z<0.2$ and magnitude-limited to $m_r = 24$ to account for optical selection biases. The predicted dwarf fractions are shown by the red curve in Figure \ref{fig:dwarffrac}, ranging from 0.05 to 0.14. Based on the current $f_{\rm dwarf}$ limit using the CHIME/FRB Outrigger sample, three additional dwarf galaxies would need to be recovered within this sample of 57 to match the predicted fraction for $\chi_{\mathrm{SFR}} = 1$. We consider whether the candidates without spectroscopic redshifts could account for this difference. For these candidates, we estimated redshifts based on the observed FRB DM following \cite{James23} and find that even adopting the upper 1$\sigma$ limit on the inferred redshift, four remain sufficiently faint, making them promising dwarf candidates.

As illustrated in Figure~\ref{fig:dwarffrac}, the measured lower limit on the CHIME/FRB Outrigger FRBs, $f_{\rm dwarf} > 0.09$, implies $\chi_{\mathrm{SFR}}>0.4$, ruling out FRBs as tracers of stellar mass alone. Comparing $f_{\rm dwarf}$ of all FRBs to the model prediction yields an agreeable match with $\chi_{\mathrm{SFR}}$ $= 1$ and aligns with an SF-weighted galaxy population, although we caution that the heterogeneous experiments contributing to the sample make the associated optical selection effects difficult to characterize. This conclusion is also consistent with previous studies based on global FRB host demographics \citep{Bhandari22, Gordon23, Sharma24}, demonstrating that the dwarf fraction alone provides a useful diagnostic of how closely FRBs trace star formation. In contrast, the dwarf fractions of CCSNe, SLSNe-I, and LGRBs all lie above the range predicted by the model. For SLSNe-I and LGRBs, this is consistent with their known preference in low-metallicity environments, thereby boosting their dwarf fractions relative to the model predictions.

Interestingly, $f_{\rm dwarf}$ for CCSNe has a sizably higher value than the predicted fraction for $\chi_{\mathrm{SFR}} = 1$ despite CCSNe being widely regarded as tracers of star formation. In other words, we would expect the CCSNe $f_{\rm dwarf}$ value to match that for the SF-weighted galaxy population. This discrepancy is likely a consequence of various factors. First, there is a spread in dwarf fractions among the CCSNe sub-classes which can be as low as $\approx0.11$ for Type Ibn, and which make up only 1-2$\%$ of all CCSNe \citep{Maeda22, Farias26}. This suggests that not all CCSN subclasses trace star formation in the same way (see Section~\ref{sec:discussion}). 

Second, host galaxy properties are intrinsically multivariate, characterized by a combination of parameters such as M$_{*}$, SFR, and $Z$ that are correlated with one another \citep{Horowicz26}. In this analysis, we consider only the simplified scenario in which the occurrence rate of a transient population depends linearly on SFR, $M_\star$, or a linear combination of the two. In other words, we do not consider the effects of metallicity on the mock galaxy population. The predicted value of $f_{\rm dwarf}$ serves to provide a reliable benchmark for distinguishing between different weighting scenarios. However, these predictions are subject to uncertainties in both the underlying galaxy population and the occurrence-rate prescriptions. As such, the completeness of \texttt{GALFRB} is limited by that of the underlying galaxy population model, which does not explicitly account for potential dependencies of FRB production on other host properties such as metallicity. Thus, a population may still broadly trace star formation while exhibiting a value of $f_{\rm dwarf}$ that differs from the SFR-only prediction. Exploring more general occurrence-rate prescriptions is beyond the scope of this work.

As alluded to earlier, these $f_{\rm dwarf}$ are contingent on the completeness of the sample and the absence of selection effects. Limitations in localization precision, host detectability, follow-up depth, or even radio properties that may affect their ability to be localized may cause faint dwarf hosts to be missed, biasing the observed value of $f_{\rm dwarf}$ low. But because we treat our measurement as a lower limit, these effects can only raise the true dwarf host fraction relative to the lower limit we placed here, resulting in possibly larger values for $\chi_{\mathrm{SFR}}$.

\input{emlines}

\section{Metallicities} \label{sec:Z}

Galaxy stellar mass and gas-phase metallicity are strongly correlated, as more massive galaxies produce and retain metals more efficiently \citep{Lequeux79,Tremonti04,Andrews13}. This mass–metallicity relation (MZR) extends to dwarf galaxies with stellar masses as small as log($M_*/M_\odot$) $\sim$ 6 \citep{Lee06,Berg12,Scholte24}. Therefore, trends in FRB host stellar masses may reflect an underlying dependence on metallicity, motivating direct measurements of the chemical abundances in FRB host galaxies. In this section, we examine specifically the gas-phase metallicity of FRB hosts, which more directly traces the present condition of star-forming regions rather than the integrated stellar metallicity of the galaxy. 

\subsection{Gas-phase Metallicity} \label{subsec:gasphaseZ}
We use oxygen abundance, 12 + log(O/H), as a proxy for the gas-phase metallicities of the FRB host galaxies using strong-line ratios (see \citealt{Zreview} for a review). Following the method outlined by \cite{Yamasaki26}, we calculate oxygen abundances using the ${\rm O_3N_2}$ diagnostic, where ${\rm O_3N_2} = \log_{10}[([\rm OIII]\lambda5007 / {\rm H}\beta) / ([\rm NII]\lambda6584 / {\rm H}\alpha)]$ \citep{PP04}. We select the ${\rm O_3N_2}$ diagnostics over other strong-line calibrations such as R$_{23}$ because the latter is double-valued for 8.0 $\lesssim$ 12 + log(O/H) $\lesssim$ 8.4 on the R$_{23}$-O/H diagram (e.g., \citealt{Pilyugin12}). This degeneracy makes $R_{23}$ unreliable for dwarf galaxies, which often lie within the metal-poor regime. To derive the gas-phase metallicities of the FRB dwarf hosts and rejected dwarf candidates, we adopt the empirical ${\rm O_3N_2}$ calibration from \citet{Curti17}. We numerically solve for the oxygen abundance, $12 + \log({\rm O/H})$, using the relation:

\begin{equation}
    \rm O_{3}N_{2} = 0.281 -4.765 x -2.268x^2
\end{equation}

\noindent where $\rm x \equiv \log({\rm O/H}) - \log({\rm O/H})_\odot$ is the oxygen abundance normalized by the solar abundance $12 + \log({\rm O/H})_\odot= 8.69$ \citep{AllendePrieto01, Asplund21}. Calibrated using the electron temperature measurements of the nebular gas producing the emission lines, this method is applicable over a range $12 + \log({\rm O/H}) = 7.6\text{--}8.85$ and is optimized for integrated galaxy spectra rather than local ${\rm H\,\text{\small II}}$ regions. The nebular emission-line fluxes and derived gas-phase metallicities are listed in Table~\ref{tab:lines}.

For a comprehensive comparison with other known FRB dwarf hosts, we additionally derive metallicities for all such hosts with sufficient emission-line fluxes to apply the ${\rm O_3N_2}$ diagnostic or the ${\rm O_3N_2}$ index itself. This includes FRBs\,20171020A, 20190417A, 20210117A, 20230708A, 20240114A, and 20250613A using emission-line fluxes reported in the literature \citep{Mahony18, Bhandari23, Bhardwaj25, Dial26, Moroianu26, Muller26}. 

For hosts with available emission-line fluxes, either derived from \texttt{Prospector} or adopted from reported measurements, we estimate metallicity uncertainties using Monte Carlo (MC) sampling of the emission-line flux uncertainties. For each emission line, we generate 1000 random draws from a Gaussian distribution centered on the flux with a standard deviation given by its uncertainty, and propagate these realizations through the metallicity diagnostics. For all FRBs considered, if any of the four emission lines lacks a reliable measurement (e.g., $[\rm NII]\lambda6584$, ${\rm H}\beta$, or both are non-detections), we adopt a $3\sigma$ upper limit based on the flux uncertainty and propagate this constraint through the metallicity diagnostics. 

Here, we discuss a few hosts with additional considerations. 
\textit{FRB\,20171020A}: As the individual emission-line fluxes are not reported, we adopt the published ${\rm O_3N_2}$ value from \citet{Mahony18} derived using the \cite{PP04} calibration. We convert the metallicity uncertainty into an equivalent ${\rm O_3N_2}$ uncertainty and propagate it through the \cite{Curti17} calibration using 1000 MC realizations.


\textit{FRB\,20190417A}: \cite{Moroianu26} identified flux-calibration issues affecting the blue wavelengths of the Gemini/GMOS spectrum, which may lead to overestimated ${\rm H}\beta$ and $[\rm OIII]\lambda5007$ fluxes. We therefore re-reduce the spectroscopic observations of FRB\,20190417A using \texttt{PypeIt} following the procedure outlined in Section~\ref{subsec:specs}. The nebular emission-line fluxes were measured using the Penalized Pixel-Fitting (\texttt{pPXF}; \citealt{Capperllari04, Cappellari17}) method, which models doublets as single blended components tied to their fixed 3:1 flux ratios. Therefore, the reported $\rm [OIII]\lambda5007$ flux and $3\sigma$ $[\rm NII]\lambda6584$ upper limit include contributions from both lines in each doublet. To recover the flux of the brighter component, we scale these measurements by a factor of 0.75. Compared to the published values, the ${\rm H}\beta$ and $[\rm OIII]\lambda5007$ fluxes from our \texttt{pPXF}-derived measurements are lower, with values of $2.5 \times 10^{-16} \rm~erg~s^{-1}~cm^{-2}$ and $1.05 \times 10^{-15} \rm~erg~s^{-1}~cm^{-2}$, respectively. 

\textit{FRB\,20210117A}: Since the VLT/FORS2 spectrum does not cover H$\beta$, we use the doublet-corrected \texttt{pPXF} measurements \citep{Bhandari23} and adopt the $N_2$ diagnostic, $\log([\rm NII]\lambda6584/{\rm H}\alpha)$ from \cite{Curti17}. This calibration is derived self-consistently with the ${\rm O_3N_2}$ calibration, and no conversion between the diagnostics is required.

\textit{FRB\,20230708A}: The emission-line fluxes reported by \cite{Muller26} were measured using \texttt{pPXF}, with the same doublet-flux correction described above already applied. We use the limit directly when computing the ${\rm O_3N_2}$ index, resulting in an upper limit on the gas-phase metallicity.

Finally, we also include a sample of SLSNe-I compiled from \cite{Lunnan14, Leloudas15, Perley16} for comparison (see Sections~\ref{subsec:Zrep} and \ref{subsec:ZPRS}). The gas-phase metallicities or upper limits are derived in the same way as described above.

\subsection{Exploring Metallicity Versus Repetition} \label{subsec:Zrep}

\begin{figure*}[t!]
    \centering
    \includegraphics[width=\textwidth]{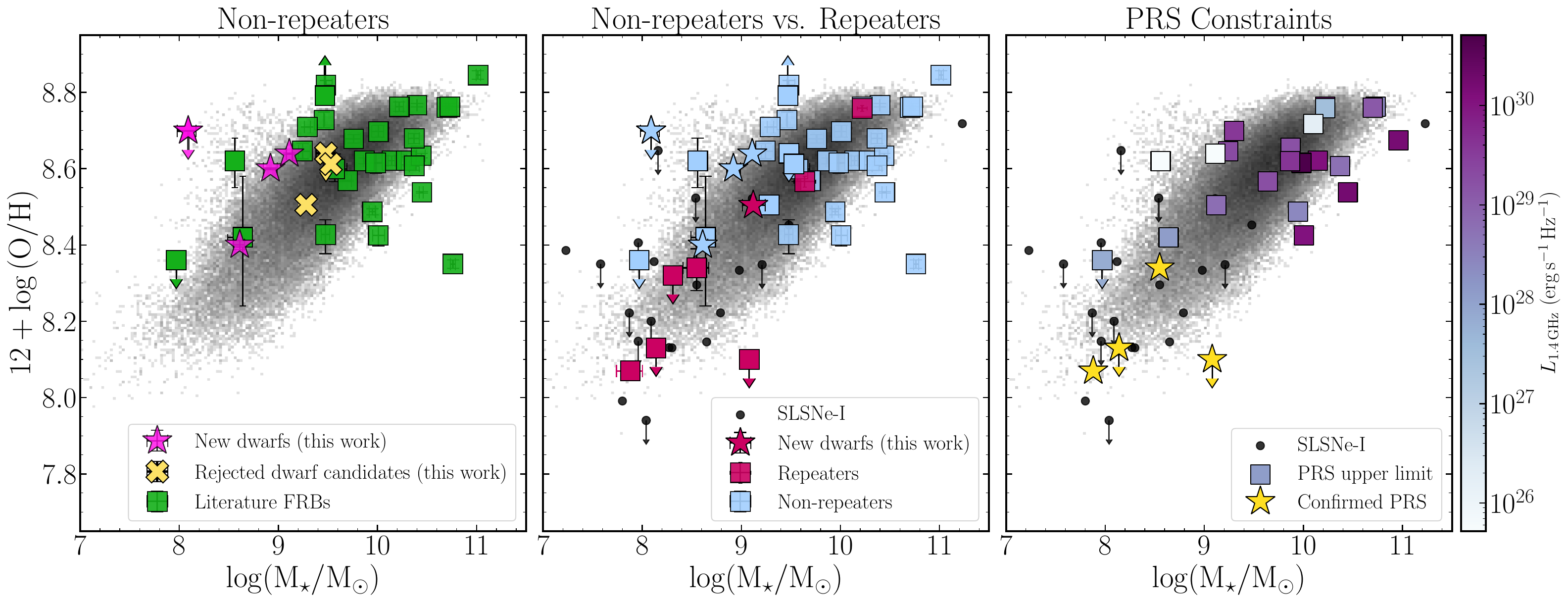}
    \caption{MZ relation comparing galaxy metallicity against stellar mass for all FRB hosts with metallicity measurements. In each panel, the gray density distribution shows a sample of field galaxies from DESI DR1 at $z<0.5$ \citep{DESIDR1}. Left: Non-repeater FRB hosts only. Magenta stars mark the new dwarf FRB hosts, yellow crosses mark the rejected dwarf candidates, and green squares mark all other non-repeater hosts from the literature at comparable redshifts \citep{Mahony18, Bhandari23, Muller26, Yamasaki26}. Middle: Same as left, but also including repeaters from this work (FRB\,20220529A) and from \cite{Bhardwaj25, Dial26, Moroianu26, Yamasaki26}; repeaters and non-repeaters are denoted as red and blue squares, respectively. The SLSNe-I hosts are shown as black circles \citep{Lunnan14, Leloudas15, Perley16}. Right: All FRBs with either upper limits on PRS counterparts or confirmed PRS detections \citep{Law22,Law24,Bhandari23,Bruni24,Dong24, Moroianu26,Muller26}. The luminosity constraints on the PRS are color-coded. The confirmed PRSs span the luminosity range of $\sim3\times10^{28-29} \rm erg~s~Hz^{-1}$ at 1.4~GHz. Non-repeater hosts, including those presented in this work, broadly occupy the same MZ locus as typical star-forming galaxies. Nearly all non-repeater hosts exhibit metallicities at or above $12 + \rm log( O/H) = 8.35$. Repeaters in dwarf galaxies have a statistically significant preference towards low metallicities (Section~\ref{subsec:Zweightings}). Moreover, repeaters in the most metal-poor hosts are also associated with a PRS (yellow stars), hinting at a possible metallicity dependence in PRS formation.}
    \label{fig:MZ}
\end{figure*}

The addition of new hosts also enables us to distinguish between repeaters and non-repeaters, particularly in the dwarf galaxy regime where known repeaters are more numerous. This partition is valuable because host galaxy studies of non-repeaters have historically been limited by the difficulty of obtaining precise localizations from their one-off bursts, in contrast to repeaters, which can be well-localized through multiple bursts. We now turn to the location of FRB hosts on the MZR. We emphasize that meaningful comparison and interpretation of the MZR require metallicities to be derived using a consistent calibration, as different strong-line calibrations based on optical emission-line ratios can produce systematic offsets of up to $\sim$1 dex in $12+\log(\mathrm{O/H})$ \citep{Kewley08}. It is therefore crucial that all metallicity measurements shown here are derived using the method described in Section~\ref{subsec:gasphaseZ}. 

In Figure~\ref{fig:MZ} (left panel), we plot the stellar masses and metallicities of the new dwarfs (pink stars) and rejected dwarf candidates (yellow crosses) presented in this work. Our sample is comprised of all apparent non-repeaters. We supplement it with other non-repeater FRB hosts, including FRBs\,20171020A, 20210117A, 20230708A, and those from \cite{Yamasaki24} within 0.01 $<z<$ 0.48 and selected based on the Baldwin–Phillips–Terlevich (BPT) diagram.

The FRB hosts are superimposed on the distribution of field galaxies with reliable redshift measurements at $z < 0.5$ from DESI DR1 \citep{DESIDR1}. This dataset comprises stellar masses 
along with fluxes of numerous emission lines, including those relevant for metallicity. To obtain reliable metallicity measurements, we adopt selection criteria similar to those outlined by \cite{Zou24}: we require signal-to-noise (S/N) $>3$ for the four emission lines needed for the oxygen abundance calculations, apply the same full-width-at-half-maximum (FWHM) cuts to reject spurious narrow detections and broad-line sources, and use the empirical \citet{Kauffman03} BPT demarcation to retain only star-forming galaxies. We further require detection of the auroral [O III]$\lambda4363$ line with a mass-dependent S/N threshold, decreasing from 3 at $\log(M_*/M_\odot)=7$ to 0 at $\log(M_*/M_\odot)=9$, to reduce contamination from compact $\mathrm{H\,II}$ regions that may be misclassified as galaxies. This selection criterion also naturally favors low-metallicity galaxies, which is useful for comparisons with low-metallicity FRB hosts.

We find that the non-repeater hosts, including those presented in this work, broadly fall within the MZ distribution of star-forming galaxies. Overall, nearly all non-repeater hosts exhibit metallicities at or above $12+\log(\mathrm{O/H})=8.3$. In the dwarf regime, FRBs\,20230708A and 20250704A have only metallicity upper limits due to the non-detection of $[\rm NII]\lambda6584$ in the former and of both H$\beta$ and $[\rm NII]\lambda6584$ in the latter (Table~\ref{tab:lines}). Nevertheless, the limit for FRB\,20230708A suggests that it may probe one of the most metal-poor environments identified to date among the non-repeater hosts. One outlier among the massive galaxies is FRB\,20230307A whose metallicity was measured by \cite{Yamasaki24}. It exhibits the lowest measured metallicity among the non-repeater hosts ($12+\log(\mathrm{O/H})=8.35$) despite its high stellar mass of $\log(M_*/M_\odot)=10.76$.

FRBs\,20250227A and 20250316A, both non-repeaters, also have measurements of their local metallicities, albeit from a different ${\rm O_3N_2}$ method (Simha et al., in prep.). For FRB\,20250316A, our global metallicity is slightly higher than the local measurement of $12+\log(\mathrm{O/H})\sim8.5$. For FRB\,20250227A, the local metallicity is only constrained to an upper limit but below than our derived global value of $12+\log(\mathrm{O/H})=8.6$.

Previously, no statistical distinction in global host properties has been found between repeaters and non-repeaters \citep{Gordon23}. With a small sample of non-repeaters now extending into the low-mass territory typically populated by repeater hosts, we have an opportunity to directly test whether these sub-populations differ in the MZ parameter space. In the middle panel of Figure~\ref{fig:MZ}, we include repeater hosts with metallicity measurements at $z<0.5$. These include dwarf hosts (FRBs\,20121102A, 20190417A, 20190520B, 20240114A, and 20250613A) and two repeaters in typical star-forming galaxies (FRBs\,20180301A and 20201124A) from \cite{Yamasaki26}, with emission-line measurements from \cite{Gordon23}.

While the repeater hosts also appear to lie along the MZR of field galaxies across the full mass range, albeit with a smaller sample size compared to non-repeaters, three of the six dwarf hosts of repeaters occupy the lowest mass and metallicity region of the MZR cloud. \textit{Five} of the repeater dwarf hosts exhibit metallicities below $12+\log(\mathrm{O/H})\sim8.35$, while three have very low metallicities of $\lesssim 8.15$ (FRBs\,20121102A, 20190520B, and 20190417A), an extremely metal-poor region in which no known non-repeater host exists at $z<0.5$. The repeater distribution is also broadly consistent with the host population of SLSNe-I, which is dominated by dwarfs with metallicities clustering around $12+\log(\mathrm{O/H})\sim8-8.3$.

To quantitatively compare the metallicity distributions of repeaters and non-repeaters across the full mass range and properly account for upper limits, we employ the log-rank test as implemented in the \texttt{lifelines} Python package \citep{Davidson-Pilon2019}. This test compares the observed and expected number of events between the repeater and non-repeater samples at each metallicity, but considers both the upper limits, or left-censored measurements, and the detections in this comparison, with equal weight for either type of data point. We reject the null hypothesis that the two distributions are drawn from the same underlying population if $p<0.01$. We find $p=0.001$, indicating that the metallicity distributions of repeaters and non-repeaters are indeed statistically distinct. Because the log-rank test does not naturally support censoring in both directions, we treat the single lower limit in our sample, FRB\,20210410D, as a conservative detection at its quoted value; excluding this source entirely does not change our conclusion.

\begin{figure*}[t!]
    \centering
    \includegraphics[width=\textwidth]{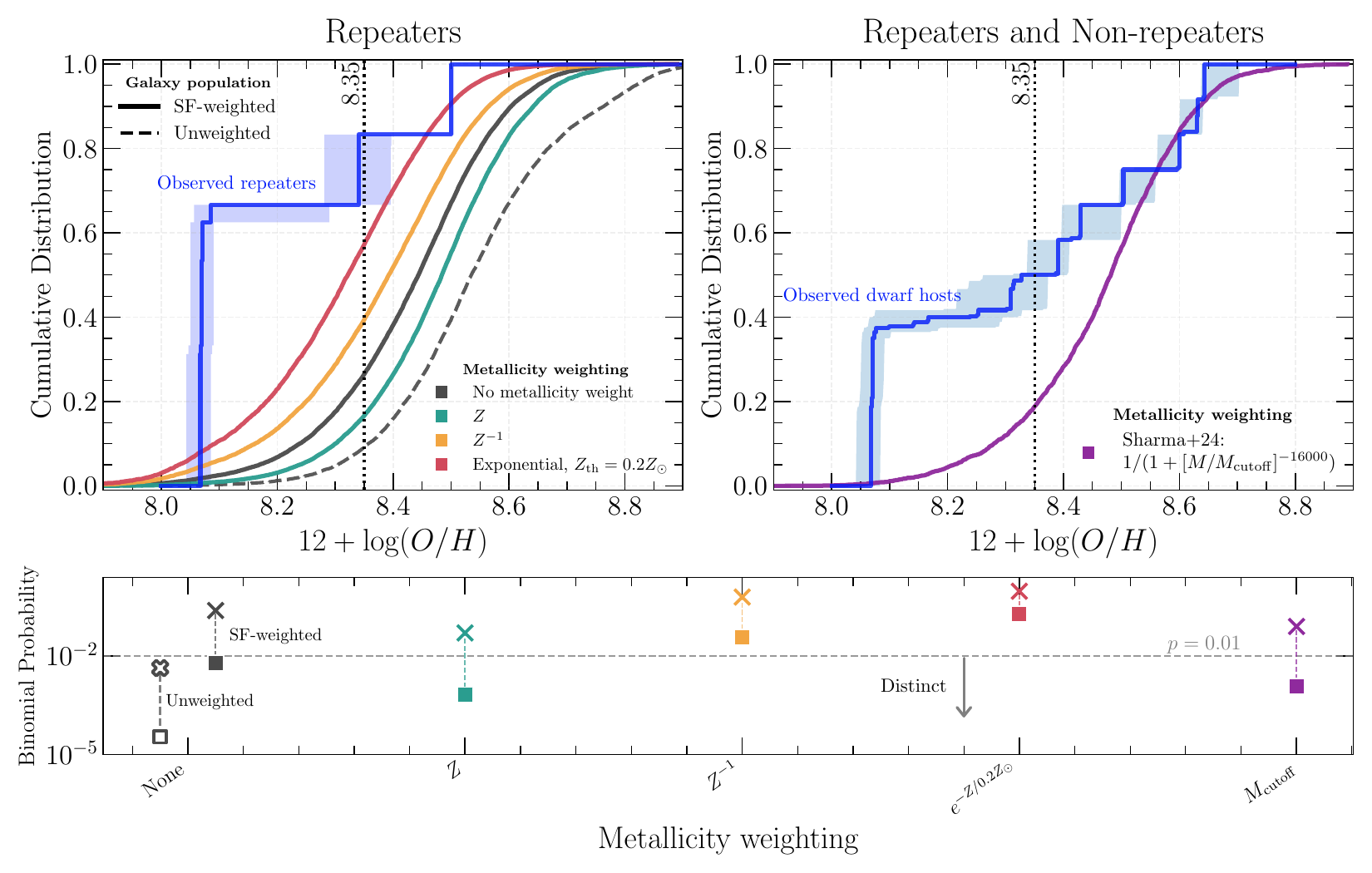}
    \caption{Host metallicity distributions for mock galaxy populations from \texttt{GALFRB} under different weighting schemes. {\it Top:} Cumulative distributions of $12+\log(\mathrm{O/H})$ for the unweighted (dashed) and SF-weighted (solid) populations, with the additional metallicity weighting schemes indicated in the legend. The stellar-mass cutoff prescription is from \cite{Sharma24}. The dark blue distribution represents the median and 68$\%$ confidence interval for the observed dwarf hosts. The dotted line marks the threshold below which most observed repeater dwarf hosts fall. {\it Bottom:} Binomial probability of drawing five hosts with $12+\log(\mathrm{O/H}) < 8.35$ and $\log(M_*/M_\odot)<9.1$ under different metallicity weightings for the repeater sample ($N=6$, squares) and full dwarf sample ($N=13$, crosses). The dashed line marks the adopted significance threshold of $p=0.01$ for the unweighted population. The change in binomial probabilities shows that a metallicity-dependent enhancement is needed to explain the observed fraction of FRB dwarf host galaxies below $12+\log(\mathrm{O/H}) = 8.35$, which are largely driven by repeaters. } 
    \label{fig:Zweighting}
\end{figure*}

\subsection{Exploring Metallicity with PRSs} \label{subsec:ZPRS}

Intriguingly, aside from the repeating FRBs\,20220529A and 20250613A, all four remaining repeaters in dwarf hosts are associated with a luminous, compact PRS. To isolate the PRS-associated FRBs within the MZ plane, we compile radio luminosities from \cite{Law22, Bruni24, Moroianu26} and highlight them as yellow stars in Figure~\ref{fig:MZ}. For ancillary data, we collect radio upper limits on PRS emission for FRBs with host metallicity measurements at $z<0.5$, where the color scale in Figure~\ref{fig:MZ} represents the PRS radio luminosity limits. For repeaters, these include FRBs\,20180301A and 20201124A\footnote{While a PRS has been proposed to be associated with FRB\,20201124A \citep{Bruni23}, the detection of resolved radio emission from obscured star formation \citep{Dong24}, together with the offset between the PRS and the FRB localization, renders this association uncertain. Given this ambiguity, we do not consider it a confirmed PRS in this work.} \citep{Law22,Dong24}, while for non-repeaters, we include FRBs\,20171020A, 20181112A, 20190714A, 20191001A, 20200430A, 20200906A, 20210117A, 20230708A, and the eleven non-repeaters discovered by the DSA-110 \citep{Law22, Bhandari23, Law24, Muller26}. The radio luminosities are scaled to a reference frequency of 1.4\,GHz assuming a power-law spectral index of $\alpha=-0.17$. This value is adopted from the median radio spectral index measured for the four confirmed PRSs \citep{Bhusare26}. 

The current upper limits on PRS emission span $\lesssim 5\times10^{30}$ to $<5\times10^{25}~\mathrm{erg~s^{-1}~Hz^{-1}}$ at 1.4~GHz. For the non-repeating FRBs\,20210117A and 20250316A, both residing in metal-rich dwarf galaxies, limits as deep as $<10^{26}~\mathrm{erg~s^{-1}~Hz^{-1}}$ at 1.4~GHz have been obtained \citep{Bhandari23, RBFLOAT}. These limits are at least three orders of magnitude below the luminosities of known FRB-associated PRSs. Nevertheless, no PRS has been detected for these FRBs thus far. Conversely, the resemblance between PRS-associated FRB hosts and the metal-poor dwarf-galaxy hosts typical of SLSNe-I as shown in Figure~\ref{fig:MZ} has already been noted in the literature \citep{Metzger17}. Our results support the notion that there may be a link between repeaters in metal-poor dwarf environments and the formation of PRSs. We stress that this connection was difficult to establish previously, given the small number of known non-repeater dwarf hosts available for comparison. 

\input{frb_radio}
\subsection{Exploring Metallicity Weightings} \label{subsec:Zweightings}

Having identified a metallicity offset between repeaters and non-repeaters, we next test whether repeaters preferentially reside in low-metallicity dwarf galaxies. We perform a series of binomial tests against a mock galaxy background at $z<0.5$ generated with \texttt{GALFRB} \citep{Loudas25}. As shown in Figure~\ref{fig:Zweighting} top panels, we draw an unweighted and a SF-weighted sample from the probability density function in ($\rm M_{*}$, SFR, z) space \citep{Leja2022}. The SF-weighted sample is further weighted according to several prescriptions for the dependence of FRB production on host $Z$, including inverse power-law and exponential forms with a characteristic metallicity threshold at $0.2Z_\odot$. We apply an optical magnitude cut of $r=24$~mag, encompassing the faintest known FRB dwarf host (FRB\,20121102A; $r=23.73$~mag), to correct for observational bias against faint dwarf galaxies. Since the framework does not inherently consider metallicity correlations with $\rm M_{*}$ and SFR as discussed in Section~\ref{subsec:dwarffrac}, we assign a $Z$ to each galaxy by employing the fundamental metallicity relation (FMR; \citealt{Sanders21}); see Appendix \ref{App.metallicity} for details. Finally, to directly address the claim of a possible metallicity cut-off for FRBs, we incorporate the $Z$-weighting prescription of \cite{Sharma24}, who find a preference for high-metallicity hosts with a stellar-mass dependence of the form 1/($1+[M/M_{\mathrm{cutoff}}]^{-\beta}$). Here, we adopt their best-fit values of $\log(M_{\rm cutoff}/M_\odot)=9.02$ and $\beta=1.6\times10^{4}$.

For comparison, in Figure~\ref{fig:Zweighting}, we also show all 13 observed dwarf hosts combining repeaters and non-repeaters. We construct their $Z$ distribution using the Kaplan–Meier estimator to account for the upper limits. We generate 1000 realizations by drawing the detected metallicities from Gaussian distributions centered on their measured values and scaled by their uncertainties, while properly accounting for our upper-limits. We show the resulting median cumulative distribution and 16th–84th percentiles in Figure~\ref{fig:Zweighting}, with the repeaters shown in the left panel and the full dwarf sample in the right panel.

Within the resulting mock galaxy populations, we compute the fraction of dwarfs with stellar masses log($M_*/M_\odot$) $<$ 9.1 that also satisfy  $12+\log(\mathrm{O/H})<8.35$. We pose the null hypothesis that the observed repeater dwarf hosts are consistent with being randomly drawn from the unweighted mock population that reflects the true field galaxy population. We adopt a significance threshold of $p<0.01$ to reject the null hypothesis. We find a probability of observing five of six hosts below the metallicity threshold is $p = 3 \times 10^{-5}$, corresponding to a $\approx 4\sigma$ significance (open black square in Figure~\ref{fig:Zweighting}). We therefore reject the null hypothesis and conclude that the concentration of repeater hosts in the metal-poor tail of the dwarf galaxy distribution is unlikely to arise by chance. 

We also consider the SF-weighted mock populations under each weighting scheme and calculate the corresponding binomial probabilities. In contrast, the binomial probability for 5 of 6 hosts below the metallicity threshold does not fall below the significance threshold of $p = 0.01 / 6$, after correcting for the family-wise error rate across six models. In particular, neither the linearly $Z$-weighted or $M_{\rm cutoff}$ weightings, shown as green and purple squares in Figure~\ref{fig:Zweighting}, can reproduce the observed concentration of repeaters in metal-poor dwarf galaxies. In contrast, the $Z^{-1}$ and exponential-$Z$ weightings (yellow and red squares), under which the hosts are randomly drawn from these more steeply metal-poor-weighted populations, provide a closer match to the observations. These results indicate that reproducing the observed concentration of repeaters in metal-poor dwarf galaxies requires FRB production to be enhanced in low-metallicity environments. Finally, the metallicity weightings explored here are intended only to demonstrate that some degree of metallicity dependence is needed to explain the observed hosts, not to identify which functional form best describes the true underlying relation.

We repeat this calculation for the full sample of 13 dwarf hosts, comprising both repeaters and non-repeaters. The resulting $p$-values, shown as crosses in the bottom panel of Figure~\ref{fig:Zweighting}, are systematically shifted to higher values relative to the repeater-only sample. The inclusion of non-repeaters brings the observed metallicity distribution into closer agreement with the models, indicating that the metallicity preference that is significant for repeaters is less pronounced when considering the full dwarf-host sample. For the $M_{\rm cutoff}$ model, \cite{Sharma24} attribute the non-detection of FRBs in dwarf galaxies to a metallicity dependence of the FRB production rate, in which the formation efficiency of the FRB source is stifled below a characteristic mass (and hence metallicity) threshold. We test this model against our full sample, calculating the binomial probability of observing 5 metal-poor ($12+\log(\mathrm{O/H})<8.35$) dwarf hosts under the null hypothesis that these hosts are randomly drawn from the $M_{\rm cutoff}$-weighted population. 
We find that we cannot reject the null hypothesis, indicating that the $M_{\rm cutoff}$ model could reproduce the observed metallicity distribution of the current dwarf-host sample. 

\subsection{Host and Burst Property Correlations}
\label{sec:propagation}

As FRB signals propagate from the source to the observer, they experience dispersion, Faraday rotation, and scattering due to the intervening ionized magnetized plasma. These propagation effects encode information about the electron density, magnetic field strength, and small-scale density fluctuations along the propagation path. In this subsection, we investigate whether these FRB properties are linked to the properties of their host galaxies across this broad range of stellar masses and metallicities in this work. For this analysis, we compile all available measurements of DM, RM, and scattering timescale ($\tau_{\rm host}$) for FRBs with published host galaxy metallicity measurements from the literature. We combine these with FRB host galaxies presented in this work, whose radio properties are summarized in Table~\ref{tab:frb_radio}, totaling 54 DM values, 37 RM values, and 24 scattering-timescale values.

\begin{figure}
    \centering
    \includegraphics[width=0.45\textwidth]{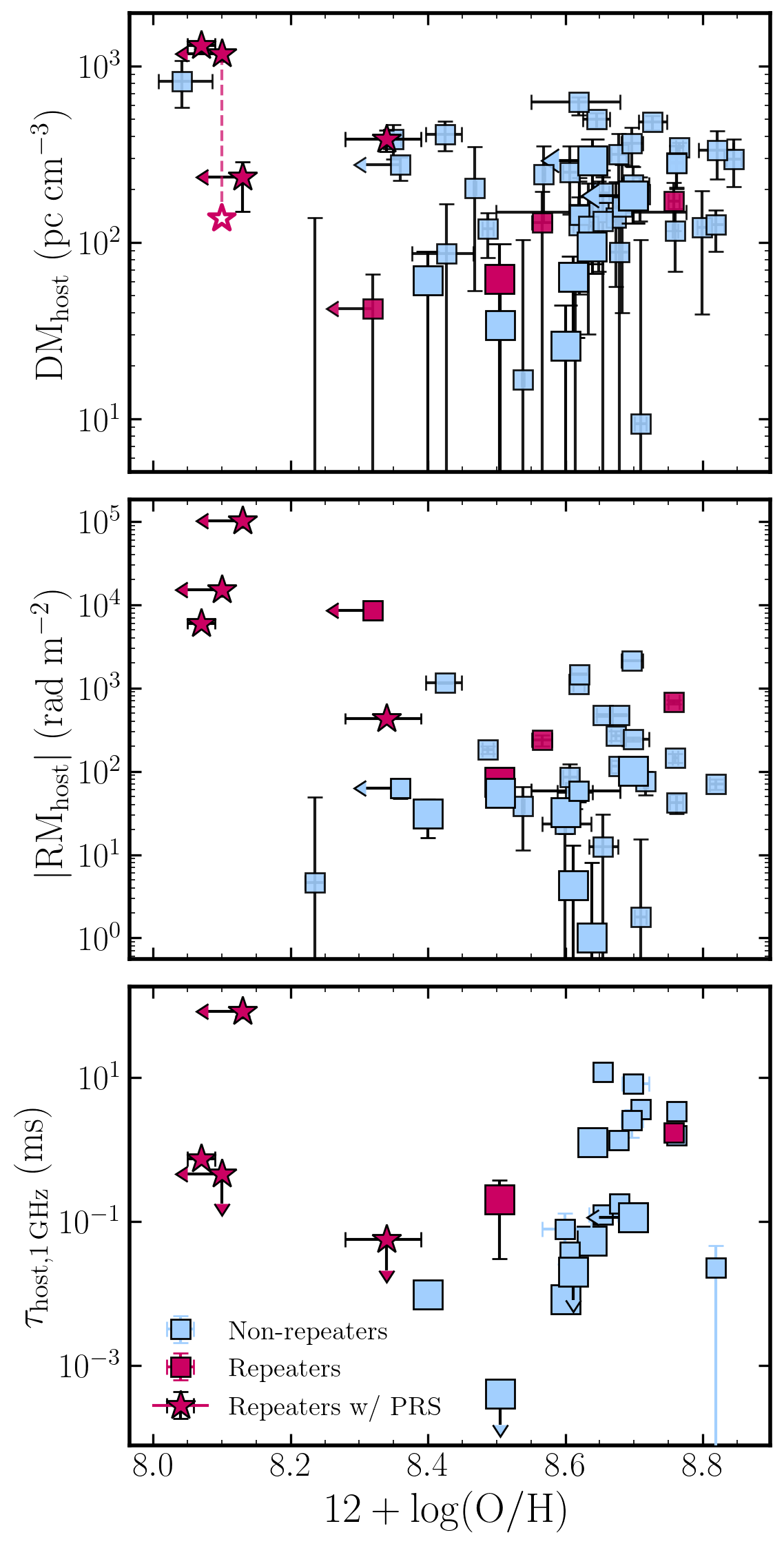}
    \caption{{DM$_\mathrm{host}$, RM$_\mathrm{host}$, and $\tau_\mathrm{host,1\,GHz}$ versus the metallicity of the host. Blue and red squares correspond to non-repeater and repeater hosts, respectively. The larger markers correspond to hosts presented in this work. Red stars mark the repeaters with an associated PRS. The dashed line and open star show the minimum $\rm DM_{\rm host}$ for FRB\,20190520B after removing contributions from foreground clusters \citep{Lee23}. All three panels share the same legend. We find no linear correlation between DM$_{\rm host}$, RM$_{\rm host}$, and metallicity, and a 2$\sigma$ positive correlation between $\tau_{\rm host,1~GHz}$ and metallicity. Excluding the PRS-associated FRBs, the $\tau_{\rm host,1,GHz}$–metallicity correlation becomes stronger (2.6$\sigma$). Based on qualitative trends, this suggests that the scattering timescale may be increasingly dominated by contributions from the host-galaxy ISM. It also implies that FRBs embedded in dense local environments such as those associated with a PRS may be biased against detection at CHIME frequencies.}}
    \label{fig:correlations}
\end{figure}

It remains an open question to what extent these propagation effects are governed by the global host properties, rather than by the local plasma environment surrounding the FRB progenitor, which is largely independent of the host. However, correlations between the global host and burst properties may be expected if these inferred propagation effects are dominated by the host ISM. For instance, higher metallicity (and more massive) galaxies have greater stellar and gas masses, which could cause more scattering by the host galaxy plasma along the sightline, resulting in a higher $\tau_\mathrm{host}$ for a given reference frequency. Here, we explore the relationship between DM$_\mathrm{host}$, RM$_\mathrm{host}$, and $\tau_\mathrm{host,1\,GHz}$ with global host galaxy metallicity in Figure\,\ref{fig:correlations}. Full details in isolating the host contribution for these burst properties are provided in Appendix~\ref{sec:burstprop_hostcalc}.

\input{correlation_tests} 

Using survival analysis, we test for monotonic correlations between metallicity and DM$_\mathrm{host}$, RM$_\mathrm{host}$, and $\tau_\mathrm{host,1\,GHz}$, considering upper limits, or left-censored, measurements. We quantify the strength of each correlation using Kendall's $\tau_{\rm k}$, a non-parametric statistic that measures the degree of concordance between two variables and is well suited to small samples and censored data. We compute the concordance index, $C$, using the survival analysis package \texttt{lifelines} \citep{Davidson-Pilon2019} and convert it to the equivalent Kendall's $\tau$ via $\tau = 2(C - 0.5)$. Since the sample includes censored measurements in both metallicity and $\tau_\mathrm{host,1\,GHz}$, we employ a generalized Kendall's $\tau$ statistic that accounts for upper limits in both parameters. For each pair of sources, we compare their relative rankings in metallicity and $\tau_\mathrm{host,1,GHz}$. Pairs for which censoring prevents the relative ordering from being established in either quantity are excluded from the calculation, affecting $<10\%$ of all possible pairings. To propagate measurement uncertainties, we perform 1000 MC realizations, drawing metallicity and propagation measurements from split-normal distributions matched to their uncertainties and repeating the calculation for each realization.

To assess the significance of the measured correlation, we perform a two-sided permutation test. The metallicity measurements, together with their censoring flags, are randomly permuted relative to DM$_\mathrm{host}$, RM$_\mathrm{host}$, and $\tau_\mathrm{host,1\,GHz}$, while the observed propagation parameters and their censoring data are preserved. The Kendall's $\tau_{\rm k}$ is recalculated for each permutation to construct the null distribution using 5000 realizations. The $p$-value is calculated as the fraction of permutations that yields $|\tau_{\rm k}|$ at least as large as the median value from the MC realizations. We report the median $\tau_{\rm k}$ and its 16th--84th percentile range, with the corresponding $p$-value in Table~\ref{tab:tau_pvalues}.

For DM$_\mathrm{host}$, we find no evidence for a correlation with metallicity for the 54 FRBs in our sample, with $\tau=-0.01^{+0.04}_{-0.05}$ and $p=0.941$. This lack of correlation is also apparent in Figure\,\ref{fig:correlations}, where most sources occupy a relatively broad distribution of DM$_\mathrm{host}$ across the metallicity range probed by our sample. The apparent high DM$_\mathrm{host}$ tail is dominated by a small number of sources, including FRB~20190417A and FRB~20190520B. For FRB~20190520B, we used the measured DM value from \citet{Connor25} without subtracting additional foreground contributions which have been shown to likely contribute significantly to the DM \citep{Lee23}. To demonstrate the impact these foreground structures, we show the minimum DM$_\mathrm{host}$ after accounting for their contributions in Figure\,\ref{fig:correlations}. For consistency, we choose to use the larger value in our statistical tests as it is derived in the same manner applied to the rest of our sample.

Recently, \citet{Athukoralalage26} extended and strengthened the claim of an anti-correlation between DM$_\mathrm{host}$ and host stellar mass reported in~\citet{Leung25} to include dwarf hosts. The low-mass end of their sample is dominated by repeaters, most of which have associated PRSs. Our sample, which extends to similarly low stellar masses but is predominantly composed of non-repeaters, provides an opportunity to test this trend with a larger sample of dwarfs. In contrast, adding our mostly non-repeating dwarf hosts to their exact sample results, we cannot recover this claimed anti-correlation between stellar mass and DM$_\mathrm{host}$, with a Pearson $r=-0.28$ and $p$-value of 0.07 (to be compared with Pearson $r=-0.5$ and $p$-value of 0.002 in \citealt{Athukoralalage26}). This suggests that the previously reported anti-correlation is sensitive to selection effects. 

One physical scenario that may reconcile these observational trends is proposed by~\citet{diggins26}, who suggest that, at fixed times $\lesssim 10$ Myr after star formation, excess DM can be efficiently produced by the harder UV continua of massive stars in low-metallicity environments. This mechanism produces strong anti-correlation between metallicity (hence, stellar mass) and host DM, but only at a fixed delay after star formation. Evidence in favor of this picture includes~\citet{li25}, who report that the host DM is correlated with specific star formation rate. 
If the delay-time distribution of non-repeating FRB sources extends to delays longer than $\sim 10$ Myr after star formation, the predicted trend would be washed out by observationally sampling both short and long delays. To reconcile~\citet{Athukoralalage26} with our findings, we speculate that the characteristic delay between star formation and FRB production is $\gtrsim 10$ Myr for our sample, which includes a larger sample of non-repeater low-mass hosts, and $\lesssim 10$ Myr for their sample (primarily repeaters at the low-mass end).

Visually in Figure\,\ref{fig:correlations}, it appears RM$_\mathrm{host}$ decreases with increasing metallicity. However, a statistical test shows there is no significant correlation in the current sample: $\tau_{\rm k}=-0.15\pm 0.03$ and $p=0.226$ for the sample size of 37 FRBs. It does appear that the sources responsible for the apparent trend in Figure\,\ref{fig:correlations} are the repeaters associated with PRSs (red stars), which occupy the high RM$_\mathrm{host}$, low metallicity region. Further, we find evidence for a $2\sigma$ positive correlation between $\tau_\mathrm{host,1\,GHz}$ and metallicity: $\tau_{\rm k}=0.31^{+0.04}_{-0.03}$ and $p=0.052$ for a sample of 24 FRBs. This correlation is not significant at the $3\sigma$ level but is in agreement with the marginal correlation reported by \citet{Glowacki_2025_PASA}. Applying a trial correction for the six tests ($p=0.01/6$) reduces the correlation significance to $1\sigma$.

The extreme and varying RM and scattering observed from FRBs associated with PRSs have been interpreted as evidence for dense, highly magnetized, and turbulent local plasma to the progenitor \citep{Michili18, Ocker22, Anna-thomas23, Moroianu26}. We can therefore assume that the local contribution to DM$_\mathrm{host}$, RM$_\mathrm{host}$, and $\tau_\mathrm{host,1\,GHz}$ dominate over the host ISM contribution for these sightlines. We, therefore, repeat our correlation analysis after excluding PRS-associated FRBs. By removing these sources, we again find no anti-correlation between RM$_\mathrm{host}$ and metallicity ($\tau_{\rm k}=0.03\pm 0.04$ and $p=0.810$ using 33 FRBs), while strengthening the correlation between $\tau_\mathrm{host,1\,GHz}$ and metallicity with a significance of $2.6\sigma$ ($\tau_{\rm k}=0.45\pm0.05$ and $p=0.008$ using 20 FRBs). However, relative to the trial-corrected significance threshold, the significance between $\tau_\mathrm{host,1,GHz}$ and metallicity drops to only $2\sigma$.

To further show the impacts of PRS systems when exploring correlations, we compare the distribution of RM$_\mathrm{host}$ between repeaters and non-repeaters in our sample. Performing Anderson-Darling rejection tests, we find that those two distributions are statistically distinct for the entire sample ($p$-value = 0.001); the difference vanishes when excluding FRB PRS sources ($p$-value = 0.09), in line with what has been reported before \citep{Pandhi_2024_ApJ}. The distinction in the population may not be between repeaters and non-repeaters \citep{Kirsten24}, but rather between FRBs embedded in extreme local environments and those for which the host-galaxy ISM provides the dominant contribution to the observed propagation effects. Whether this represents a strict dichotomy, for example between FRBs embedded in dense nebulae and those without such environments, or instead reflects a continuum of local environmental densities remains unclear (e.g. \citealt{Yang_2020_ApJ}).

This distinction is also important when considering the known CHIME/FRB selection effects against large scattering timescales in the CHIME band (400-800~MHz;\citealt{Cat1, McGregor_2026_arXiv}). If the correlation between scattering and metallicity  persists in a larger sample, then FRBs whose propagation effects are dominated by the host ISM should \textit{not} be subject to a selection bias against low-metallicity environments. Instead, our results indicate that scattering selection bias would preferentially miss FRBs in higher metallicity hosts. Additionally, for FRBs whose  host propagation contributions are predominantly from the local environment such as a PRS, CHIME/FRB will be biased against FRBs embedded in extremely dense environments, where the scattering timescale is very high \citep{Cat1,Merryfield23, McGregor_2026_arXiv}. 

Notably, our results have implications for the observed population of PRS-associated FRBs. As a simple illustration, suppose that the PRS-associated FRBs all have similar local densities and therefore similar local scattering timescale contributions. The total scattering would depend on both this local contribution and the additional scattering introduced from the ISM, which potentially increases with metallicity (Figure\,\ref{fig:correlations}). Thus, PRS-associated FRBs in higher-metallicity hosts could be preferentially missed because their local and host-ISM scattering contributions combine to produce highly scattered bursts. If the scattering--metallicity correlation is confirmed with a larger sample, this selection effect will need to be considered when interpreting the observed PRS fraction and its dependence on their host galaxies.

\section{Implications From a Population of FRB Dwarf Hosts} \label{sec:discussion}

We now discuss the implications of our results in the context of possible progenitor pathways and observations of other transients and their host environments. As a population, we find that FRBs occur across the full observable range of galaxy stellar masses and metallicities. Coupled with previous work on the global demographics of FRBs \citep{Gordon23}, this result implies that, among FRBs originating from star-forming galaxies, no specific environmental conditions on global stellar-population scales are required to produce an observable FRB. Since the majority of magnetars, including those which may produce FRBs, likely form from the core collapse of massive stars (i.e., \citealt{HuZhang2026, Pardo-Araujo2026}), this suggests that these magnetars can come from a wide range of massive star types, which in turn can trace different environments. Indeed, there is a great diversity in the types of massive stars that undergo core-collapse, i.e., zero-age main sequence (ZAMS) masses of $\sim 8-17\,M_{\odot}$ for normal CCSNe to $\gtrsim 40\,M_{\odot}$ for SLSNe and LGRB progenitors \citep{Heger2003,GalYam2017}. These progenitors, in turn, give rise to a variety of supernovae and occasionally, the specific environmental properties conducive to their formation.

Case in point, the high prevalence of SLSNe-I and LGRBs in dwarf galaxies ($\approx 70-80\%$; \citealt{Fruchter06,Lunnan14, Perley16, Taggart21}) and their strong preference for low-metallicity galaxies relative to the field \citep{Levesque10,GrahamFruchter13,Leloudas15, Schulze18, Schulze21} suggest that these environmental conditions enhance the production efficiency of their progenitors. Their unique environments provide some of the \textit{strongest} evidence that they originate from particularly young and massive stars (ZAMS masses of $M_{\rm ZAMS}\gtrsim 40\,M_{\odot}$). By the same token, we have shown that there is a higher prevalence of repeater hosts at the lowest stellar mass and metallicity end of the galaxy distribution ($12+\log(\mathrm{O/H})\lesssim 8.2$). Moreover, all known repeaters with associated PRS counterparts reside in metal-poor environments. Since the magnetars invoked to power SLSNe-I need rapid spins to produce their extreme luminosities \citep{Kasen10}, this suggests a plausible connection between rapidly rotating magnetars and the presence of FRB repetition or a PRS.

One possibility is that metallicity influences the evolution of the massive star progenitors, and in particular those that give rise to PRS counterparts. In more metal-enriched environments, stronger line-driven stellar winds lead to more efficient stripping of the hydrogen and helium envelopes prior to core collapse, resulting in greater mass and angular momentum loss \citep{Estrom12}. If FRBs are powered by a compact object such as a magnetar, the remnant formed in metal-rich environments may retain less angular momentum, producing a more slowly rotating central engine with a smaller rotational energy reservoir and less efficient magnetic-field amplification.
Conversely, lower-metallicity progenitors experience weaker stellar winds \citep{Maeder01}, allowing them to retain more angular momentum and potentially form young, rapidly-spinning magnetars with large energy budgets \citep{Thompson04, Song23}. This rapid rotation enables efficient convective and magneto-rotational dynamos, amplifying the magnetic field and increasing the energy reservoir available to produce detectable FRBs \citep{Duncan92, Mosta15, Raynaud20}. Indeed, observations indicate that extreme rotators such as OBe stars preferentially occur in metal-poor dwarf galaxies \citep{Schootemeijer22}. If these stellar progenitors produce rapidly rotating compact objects, they could more readily power both luminous repeaters and long-lived bright PRSs \citep{Kashiyama17, Metzger17, Margalit18}, consistent with the observed trend in the MZ plane shown in Figure~\ref{fig:MZ} and with previous results in the literature \citep{Moroianu26}. 

On the other hand, at $\log(M_\star/M_{\odot})\lesssim9$, the hosts of non-repeaters appear to largely lie above $12+\log(\mathrm{O/H})\gtrsim8.4$, a trend that extends across the full stellar-mass range. Given that non-repeaters constitute the bulk of the observed FRB population, this could naturally be explained if the dominant FRB progenitor channel involves magnetars formed without extreme birth spins \citep{HuZhang2026}, and therefore does not require the low-metallicity environments that may favor the formation of rapidly rotating magnetars. Drawing comparisons from the Milky Way, the Galactic magnetars are generally inferred to originate from relatively ordinary massive stars with ZAMS masses of $\lesssim20~M_{\odot}$ \citep{Davies09}. In fact, detailed studies of SGR~1935$+$2154 associated with FRB\,20200428D \citep{Bochenek2020, CHIME2020-magnetar}, its SN remnant, and surrounding environment find no evidence for a very massive progenitor and indicate that its associated supernova was not particularly energetic \citep{Zhou2020, He26}. Within the CCSNe population, stripped-envelope SNe (Type Ib/c) are also typically found in more metal-rich environments than Type II SNe \citep{VandenBergh97, Anderson10}, highlighting how differences in the massive star population that give rise to magnetars can manifest as distinct host galaxy trends. 

A caveat to this interpretation is that it may also be affected by selection effects. Given the strong correlation between galaxy stellar mass and metallicity (Section~\ref{sec:Z}), if non-repeaters with PRSs exist across a broader range of host metallicities and have similarly dense local environments, those in higher-metallicity galaxies may be preferentially missed by CHIME if an underlying scattering-metallicity correlation is uncovered in a larger sample (as described in Section\,\ref{sec:propagation}). This could therefore produce an artificial truncation in the observed metallicity distribution, making PRS-associated FRBs appear preferentially associated with low-metallicity hosts.

We also consider the possibility of a metallicity-dependent suppression in FRB formation, as proposed by \cite{Sharma24}, with a characteristic cutoff at $12+\log({\rm O/H})=8.08$. As discussed in Section~\ref{subsec:Zweightings}, this prescription is statistically consistent with the metallicity distribution of the observed FRB dwarf hosts, yielding $p = 7 \times 10^{-2}$. Thus, given the current sample size, we cannot rule out a model in which a metallicity cutoff near this threshold suppresses FRB formation in low-metallicity galaxies. However, when considering only the repeaters, we obtain $p = 1 \times 10^{-3}$, indicating that this model is inconsistent with the observed concentration of repeaters in metal-poor dwarf galaxies.

Throughout Section~\ref{sec:Z}, we have separated the FRB population into repeaters and non-repeaters when investigating their host metallicities and correlations with various observed burst properties. We note, however, that some of the apparently non-repeating FRBs in our sample may eventually be found to repeat with future observations. To further assess this distinction, we show the burst morphologies of the six new non-repeaters from this work in Figure~\ref{fig:waterfalls}. They generally exhibit broad bandwidths and show no clear evidence of downward-drifting structures typically associated with repeaters (e.g., \citealt{Hessels_2019_ApJL, Pleunis21}). For the bright and nearby FRB\,20250316A, CHIME has not detected any additional bursts consistent in position and DM over $\approx$270 hours of exposure time spanning six years. Furthermore, a Poisson rate above $1.5\times10^{-2}$ bursts hr$^{-1}$ has been ruled out at 99.7$\%$ confidence \citep{RBFLOAT}. While continued monitoring will ultimately be required to better explore whether these sources are intrinsically non-repeating, their burst morphologies and the existing observational limits provide little indication of repetition at present. We finally recognize that the distinction between repeaters and non-repeaters is not definitive, as they may be reconciled within a single underlying population with a broad distribution of repetition rates \citep{James23,Yamasaki24,Beniamini25, Cook26}.

\section{Conclusions \& Future Outlook} \label{conclusion}
We performed a systematic search to identify and characterize a new population of FRB dwarf-galaxy hosts using a sample of 78 FRBs localized by CHIME/FRB Outriggers. We performed SED fitting to infer their host stellar masses, compared their stellar mass distributions with those of other known transients, and placed a lower limit on the fraction of FRBs in dwarf galaxies. We further measured and investigated the global host galaxy metallicities in relation to FRB repetition and PRS associations, as well as potential correlations between global host properties and FRB burst properties that may provide insight into propagation effects along the line of sight. Our main conclusions are as follows:

\begin{itemize}
    \item We identified \ndwarfs~new dwarf-galaxy FRB hosts, with stellar masses of $\log(M_{\ast}/M_{\odot}) \leq 9.1$, as well as three dwarf galaxy candidates with higher stellar masses of $\log(M_{\ast}/M_{\odot}) \approx 9.3$–$9.5$ that we ultimately rejected as dwarf hosts in our sample.

    \item We place a lower limit on the FRB dwarf fraction, $f_{\rm dwarf} > 0.09_{-0.03}^{+0.04}$, using the CHIME/FRB Outrigger sample at $z<0.2$. Both this limit and the fraction obtained when including literature hosts are consistent with the observed dwarf fractions of SLSNe-I, LGRBs, CCSNe, and Type Ia SNe. By comparing our results with mock galaxy populations weighted by stellar mass ($\chi_{\rm SFR}=0$), star-formation rate ($\chi_{\rm SFR}=1$), and intermediate weightings, our lower limit on FRB $f_{\rm dwarf}$ implies $\chi_{\rm SFR} > 0.4$, ruling out progenitors tracing an old stellar population alone.

    \item Separating the FRB host sample into repeaters and non-repeaters, we find that the metallicities of non-repeater hosts are broadly consistent with the stellar mass-metallicity distribution of star-forming galaxies, but are largely restricted to $12+\log(\mathrm{O/H}) \gtrsim 8.35$. In contrast, repeater hosts preferentially occupy lower-metallicity environments, in which most of the dwarf hosts exhibit $12+\log(\mathrm{O/H}) < 8.35$, indicating a statistically significant preference for metal-poor dwarf galaxies. We further find that the metallicity distributions of repeaters and non-repeaters are statistically distinct with $p=0.001$. Moreover, almost all PRS-associated repeaters reside in the most metal-poor dwarf galaxies with $12+\log(\mathrm{O/H}) < 8.2$, suggesting a connection between these environments and PRS production.

    \item The global metallicities of FRB hosts show no significant correlation with $\mathrm{DM}_{\rm host}$, $\mathrm{RM}_{\rm host}$, or $\tau_{\rm host,1~GHz}$. Removing PRS-associated repeaters yields a tentative correlation (2$\sigma$ after trial correction) between $\tau_{\rm host,1,GHz}$ and metallicity, but it remains insignificant at the $3\sigma$ level. This could, if demonstrated to be significant with a larger sample, indicate that the scattering timescale is dominated by the host-galaxy ISM rather than the local FRB environment for most FRBs.

    \item The tentative correlation between scattering and host-galaxy metallicity highlights the role of radio selection effects that future studies may need to consider, particularly at CHIME frequencies (400-800~MHz). This has important implications for the observed population of PRS-associated FRBs, as such sources may be underrepresented in high-metallicity environments, where scattering from the host-galaxy ISM combined with large local scattering could suppress FRB detections.

    \item The distinct locations of repeaters and non-repeaters in the MZ plane may be explained by a metallicity dependence in the evolution of massive stars that give rise to FRB progenitors. In particular, lower-metallicity environments may favor more rapidly rotating progenitors, which could enhance magnetic-field amplification and provide sufficient energy to power repeaters and their associated PRSs. In contrast, the majority of non-repeaters, which occupy higher-metallicity environments, may arise from magnetars with weaker birth spins formed through the evolution of more ordinary massive stars with ZAMS masses $\lesssim20~M_{\odot}$ \citep{HuZhang2026}.

\end{itemize}

Throughout this work, we have discussed several observational biases that may hinder the identification of faint FRB dwarf hosts. Looking to the future, one remedy for recovering the missing dwarf FRB hosts is to leverage next-generation all-sky optical surveys, including Rubin/LSST, Roman, and Euclid \citep{LSST, Roman, Euclid}. Rubin, for example, is imaging 18,000 $\rm deg^2$ of the Southern sky, reaching 5$\sigma$ depths (AB mag) of $m_r\approx 24$~mag for single exposures and $m_r\approx 27$~mag for the 10-year co-added stack. This increased depth will certainly enable the discovery of many more dwarf galaxies. Such surveys will also alleviate the need for dedicated optical follow-up and are well suited to expanding the volume over which we are sensitive to dwarf FRB hosts, particularly improving observational completeness for nearby hosts.

In terms of radio selection effects, upcoming radio surveys, e.g. the Canadian Hydrogen Observatory and Radio-transient Detector (CHORD; \citealt{CHORD}) and the DSA \citep{DSA-2000, Connor26}, will discover tens of thousands of FRBs up to higher radio frequencies (1.5\,GHz and 2\,GHz, respectively) than CHIME. These higher frequencies will aid in reducing the impact of scattering on the detectability of FRBs (since $\tau\propto\nu^{-4}$). This will provide an opportunity to explore whether FRB-PRS systems are preferentially associated with low-metallicity hosts or whether the apparent concentration in these environments currently is produced by radio selection effects. If PRSs are intrinsically associated with low-metallicity environments, and contribute high local scattering, the reduced scattering bias of these higher-frequency surveys should enable the detection of a larger population of PRS-associated FRBs.

\begin{acknowledgments}
We are grateful to Alexa Gordon, Joel Leja, Yijia Li, Anya Nugent, and Shotaro Yamasaki for the extremely helpful discussions. Y.D. is grateful for the support from the National Science Foundation (NSF) Graduate Research Fellowship under grant No. DGE-2234667. W.F. gratefully acknowledges support by the NSF under grant no. AST-2206494 and CAREER grant No. AST-2047919, the David and Lucile Packard Foundation, the Alfred P. Sloan Foundation, and the Research Corporation for Science Advancement through Cottrell Scholar Award \#28284. K.N. acknowledges support by NASA through the NASA Hubble Fellowship grant \# HST-HF2-51582.001-A awarded by the Space Telescope Science Institute, which is operated by the Association of Universities for Research in Astronomy, Incorporated, under NASA contract NAS5-26555. A.M.C. is a Banting Postdoctoral Fellow. C.L. acknowledges support from the Miller Institute for Basic Research at UC Berkeley. S.S. acknowledges support from the Brinson Foundation as a NU-UC joint fellow. 

A.P.C. is a Canadian SKA Scientist and is funded by the Government of Canada / est financé par le gouvernement du Canada. V.M.K. holds the Lorne Trottier Chair in Astrophysics \& Cosmology, a Distinguished James McGill Professorship, and receives support from an NSERC Discovery grant (RGPIN 228738-13). K.W.M. is supported by NSF Grant No. 2510771. A.B.P.~acknowledges support by NASA through the NASA Hubble Fellowship grant \mbox{HST-HF2-51584.001-A} awarded by the Space Telescope Science Institute, which is operated by the Association of Universities for Research in Astronomy, Inc., under NASA contract \mbox{NAS5-26555}. A.B.P.~also acknowledges prior support from a Banting Fellowship, a McGill Space Institute~(MSI) Fellowship, and a Fonds de Recherche du \mbox{Qu\'ebec -- Nature} et Technologies~(FRQNT) Postdoctoral Fellowship. P.S. acknowledges the support of an NSERC Discovery Grant (RGPIN-2024-06266). M.W.S. acknowledges support from the French government under the France 2030 investment plan, as part of the Initiative d’Excellence d’Aix-Marseille Université - AMIDEX (AMX-23-CEI- 088).

The Fast and Fortunate for FRB Follow-up team acknowledges support from NSF grants AST-1911140, AST-1910471, and AST-2206490. CHIME operations are funded by a grant from the NSERC Alliance Program and by support from McGill University, University of British Columbia, and University of Toronto. CHIME/FRB Outriggers are funded by a grant from the Gordon \& Betty Moore Foundation. We are grateful to Robert Kirshner for early support and encouragement of the CHIME/FRB Outriggers Project, and to Dusan Pejakovic of the Moore Foundation for continued support. The CHIME/FRB Project was funded by a grant from the CFI 2015 Innovation Fund (Project 33213) and by contributions from the provinces of British Columbia and Québec, and by the Dunlap Institute for Astronomy and Astrophysics at the University of Toronto. Additional support was provided by the Canadian Institute for Advanced Research (CIFAR), the Trottier Space Institute at McGill University, and the University of British Columbia. 

The AstroFlash research group at McGill University, University of Amsterdam, ASTRON, and JIVE is supported by: a Canada Excellence Research Chair in Transient Astrophysics (CERC-2022-00009); an Advanced Grant from the European Research Council (ERC) under the European Union's Horizon 2020 research and innovation programme (`EuroFlash'; Grant agreement No. 101098079); an NWO-Vici grant (`AstroFlash'; VI.C.192.045); an NSERC Discovery Grant (RGPIN-2025-06681); an ERC Starting Grant (`EnviroFlash'; Grant agreement No. 101223057); and an NWO-Veni grant (VI.Veni.222.295).

This research has made use of the Keck Observatory Archive (KOA), which is operated by the W. M. Keck Observatory and the NASA Exoplanet Science Institute (NExScI), under contract with the National Aeronautics and Space Administration. This work made use of the legacystamps package (https://github.com/tikk3r/legacystamps).

\end{acknowledgments}

%
\facilities{CTIO:2MASS, DESI, Gemini:Gillett (GMOS), Gemini:South (GMOS), Keck:I (LRIS), MMT(MMIRS), PS1, WISE}



\software{
{\tt astropy} \citep{Astropy13, Astropy18, Astropy22},
{\tt dynesty} \citep{Speagle20},
{\tt FFFF-PZ} \citep{Coulter2022, Coulter2023},
{\tt FRBs/FRB} \citep{FRBrepo},
{\tt lifelines} \citep{Davidson-Pilon2019},
{\tt marz} \citep{Marz},
{\tt matplotlib} \citep{Matplotlib},
{\tt numpy} \citep{numpy}, 
{\tt pandas} \citep{reback2020pandas}, 
{\tt photutils} \citep{photutils},
{\tt POTPyRI\footnote{\url{https://github.com/CIERA-Transients/POTPyRI}}},
{\tt Prospector} \citep{Johnson21},
{\tt PypeIt} \citep{pypeit:joss_arXiv, pypeit:zenodo},
{\tt python-fsps} \citep{Conroy09, Conroy10},
{\tt SAOImageDS9} \citep{DS9},
{\tt scipy} \citep{Scipy}, 
{\tt sedpy} \citep{sedpy}
}


\clearpage
\appendix
\restartappendixnumbering
\section{FRB waterfall plots}
De-dispersed dynamic spectra of the new FRB host galaxies presented in this work with previously unpublished baseband data. FRBs\,20250206A, 20250227A, and 20250704A are confirmed dwarf hosts, while FRBs\,20250410E, 20250507A, and 20250515A were rejected as dwarfs based on their stellar masses. For the remaining two, FRB\,20250316A and the repeating FRB\,20220529A, dynamic spectra have already been published by \citet{RBFLOAT} and \citet{Pandhi26}, respectively.

\begin{figure}[H]
    \centering
    \includegraphics[width=\linewidth]{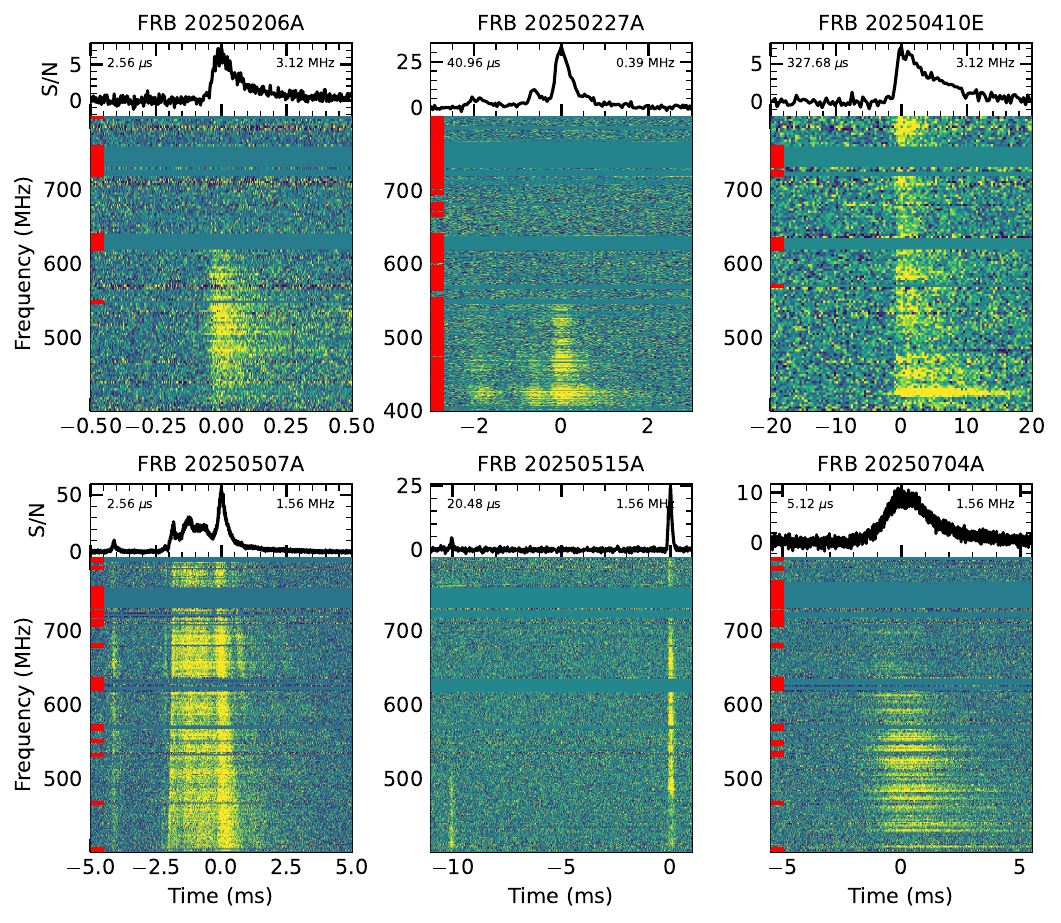}
    \caption{De-dispersed dynamic spectra (bottom sub-panels) and frequency-averaged profiles (top sub-panels) for each FRB, dedispersed to the best-fitting DM reported in Table\,\ref{tab:basics}. The time and frequency resolution used for plotting is indicated in the upper left and right of each panel, respectively. Frequency channels contaminated by radio frequency interference have been excised and are marked by the red indicator along the left edge of each dynamic spectrum.}
    \label{fig:waterfalls}
\end{figure}

\section{Photometry}
Here, we list the photometric observations for all new FRB hosts presented in this work.

\input{photometry}

\section{SED Fitting Details} \label{appendix:SEDfits}

We initiate all of our \texttt{Prospector} fits with a \cite{Kroupa01} initial mass function and \cite{KC13} dust attenuation curve. To ensure physically realistic total mass formed ($M_F$) and stellar metallicity ($Z_*$) posterior distributions, we impose a prior that approximates a Gaussian scatter around the \cite{Gallazzi05} $M-Z_*$ relation. We apply a continuum normalization model to match the spectrum to the photometry, a spectral smoothing model, a pixel outlier model to marginalize over poorly modeled noise, and a jitter model that multiplicatively inflates the uncertainties of all spectroscopic pixels, as needed, to achieve a good fit (see Appendix D of \citealt{Johnson21} for details). We fit the spectral emission lines using a nebular marginalization template based on \texttt{Cloudy} photoionization model \citep{Byler17} and set both the gas-phase metallicity ($Z_{gas}$) and gas ionization ($U_{gas}$) parameter as free. Nebular marginalization is not included for FRB\,20250410E, as the fit failed during sampling due to numerically unstable matrix inversion.

We impose a $5\%$ error floor on all photometric fluxes to prevent overfitting to any single measurement. For any host with WISE photometry, we incorporate both dust emission and active galactic nuclei (AGN) templates to properly model the mid-IR regime. We employ the three-component dust emission model from \cite{DraineLi07}, in which dust grains are modeled as polycyclic aromatic hydrocarbons (PAHs). We set only $q_{pah}$, the mass fraction of dust in PAH form, as a free parameter as the WISE photometry is not sensitive to either the minimum radiation strength ($U_{min}$) or the fraction of dust in high radiation fields ($\gamma$). We use the two-component AGN model from \cite{Nenkova08} that describes the mid-IR optical depth ($\tau_{AGN}$) and the total AGN luminosity ($f_{AGN}$), and set both parameters as free.

In general, we adopt eight age bins, except when only one or two photometric points are available, as in the cases of FRBs\,20250410E and 20250704A. For these fits, we reduce to five age bins, as \cite{Leja19} showed that SED fits are largely invariant to the number of age bins provided that they exceed four. The first two bins are fixed at 0–30 Myr and 30–100 Myr, with the youngest bin being the narrowest, and the maximum edge of the oldest bin fixed at the age of the Universe at the redshift of each host galaxy. All remaining age bins are spaced equally in logarithmic time.

\begin{figure}[H]
    \renewcommand{\thefigure}{C1}
    \centering
    \vspace*{-3cm}
    \includegraphics[
        width=\textwidth,
        height=\textheight,
        keepaspectratio
    ]{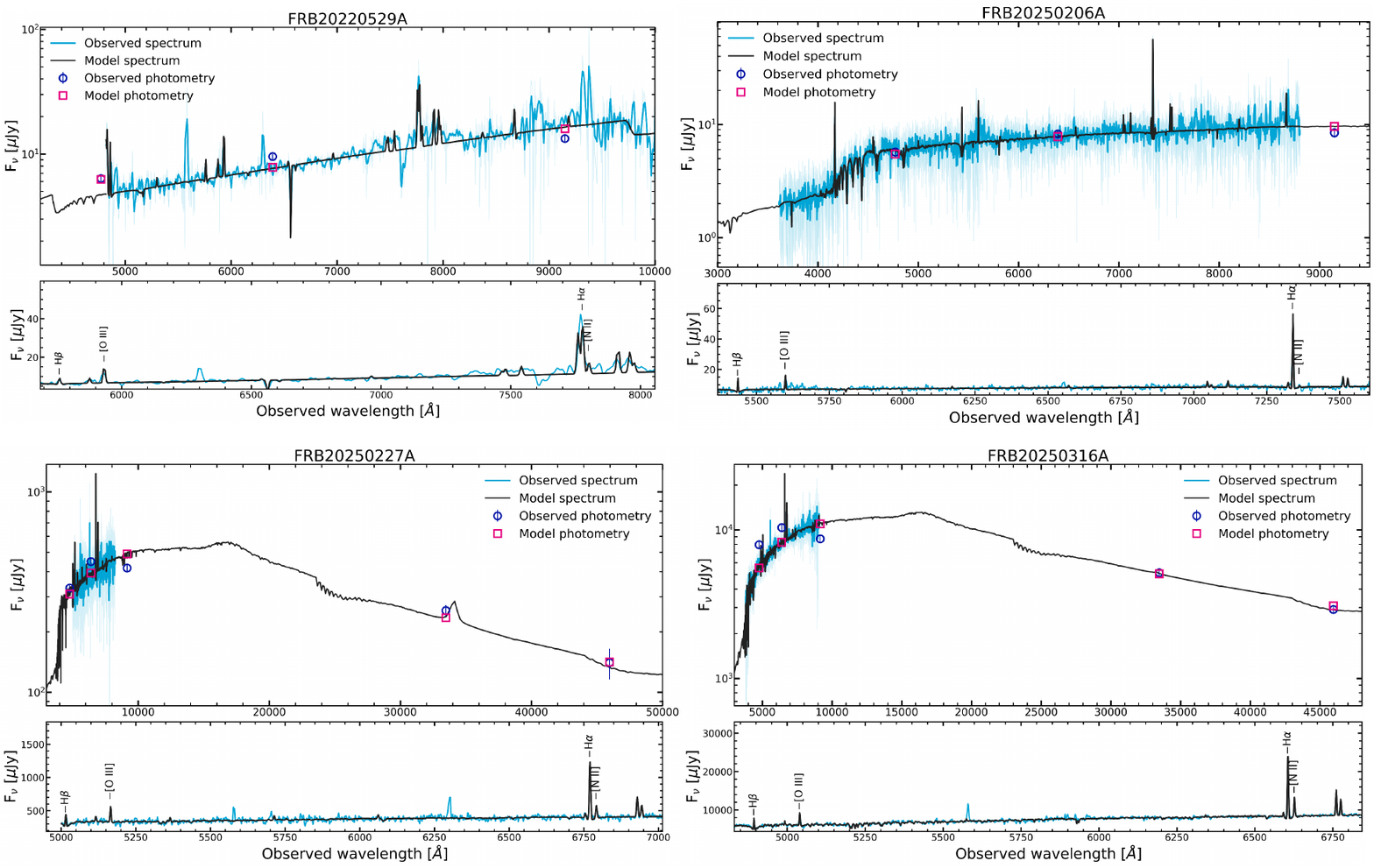}
    \vspace*{-1.2cm}
    \caption{The broad-band observed photometry (blue circles), observed spectrum (blue line), best-fit model spectrum (black line), and model photometry (pink squares) for our dwarf hosts and rejected dwarf candidates from the non-parametric model fitting using \texttt{Prospector}.}
    \label{fig:seds}
\end{figure}

\begin{figure}[H]
    \centering
    \includegraphics[
        width=\textwidth,
        height=\textheight,
        keepaspectratio
    ]{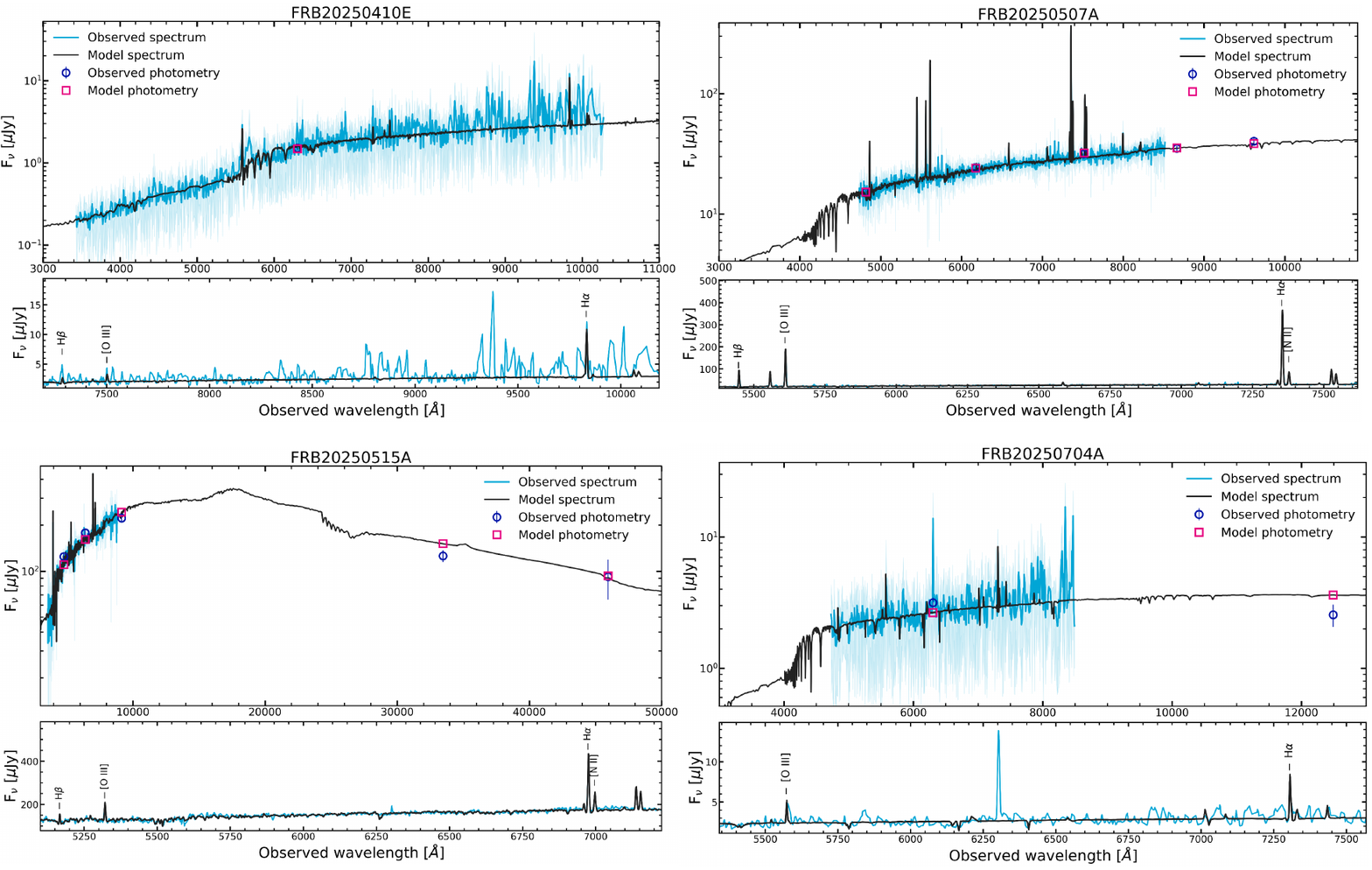}
    \vspace{-1.2cm}

    \textbf{Figure C1 (cont.).} The broad-band observed photometry (blue circles), observed spectrum (blue line), best-fit model spectrum (black line), and model photometry (pink squares) for our dwarf hosts and rejected dwarf candidates from the non-parametric model fitting using \texttt{Prospector}.
\end{figure}


\section{Assigning metallicity to the \texttt{GALFRB} galaxy samples}
\label{App.metallicity}

\texttt{GALFRB} generates mock galaxy samples with $\rm (M_\star, SFR, z)$ information. We added metallicity values to each mock galaxy in a postprocessing step following \cite{Yamasaki26}, which exploits the fundamental metallicity relation (FMR) calibrated on SDSS data \citep{Sanders21} in which
\begin{equation}
    \rm 12 + \log(O/H) = 8.80 + 0.188y - 0.220y^2 - 0.0531y^3
\end{equation}
and $\rm y= \mu_\alpha(M_\star, SFR) - 10$ is the metallicity parameter \citep{Mannucci2010}. Here,
\begin{equation}
    \rm \mu_\alpha = \log M_\star + \log M_{offset} - \alpha \times \log SFR.
\end{equation}
We set $\alpha=0.60$ to minimize the scatter in $\rm \mu_\alpha - Z$ space \citep{Sanders21} and a stellar mass offset $\rm \log(M_{offset}) = - 0.20$ that corrects for the systematic bias between the SDSS galaxies modeled with parametric SFHs and the \texttt{GALFRB} samples that rely on galaxy data fit with non-parametric SFHs (see discussion in \citealt{Leja2022} and \citealt{Yamasaki26}).
Finally, to account for the observed intrinsic scatter in metallicity, we included a scatter of $\rm 0.05~dex$ in $\rm \log Z$.

\section{Host DM, RM, and Scattering Timescale Estimations} \label{sec:burstprop_hostcalc}

The observed DM, RM, and scattering timescale each contains contributions from every magneto-ionized medium along the line of sight, including the Milky Way (MW) interstellar medium (ISM) and halo, intergalactic medium (IGM), and the FRB host galaxy. For DM and RM host contributions, we perform a MC decomposition following \citet{Pandhi_2025_ApJ}, calculating $\rm DM_{\rm host} = (DM_{\rm obs} - DM_{\rm MW,disk} - DM_{\rm MW,halo} - DM_{\rm IGM})\times (1+z)$ and $\rm RM_{\rm host} = (\rm RM_{\rm obs} - RM_{\rm MW} - RM_{\rm IGM}) \times (1+z)^2$ for 10,000 realizations. In each realization, we sample from the following distributions: 
\begin{itemize}
    \item $\rm DM_{\rm obs}$ and $\rm RM_{\rm obs}$ are taken from the measured values. In this work, we compute the structure-maximizing DM using {\tt DM\_phase}\footnote{\url{https://github.com/danielemichilli/DM_phase}}, which we refine using {\tt fitburst} \citep{Fonseca_2024_ApJS} to simultaneously fit for DM and scattering, and the RM using the CHIME/FRB polarization pipeline described by \citet{Mckinven21}. The measured quantities are reported in Table\,\ref{tab:frb_radio}. We compile literature values from  \cite{Chatterjee17, Banister19, Prochaska19, Hilmarsson_2021_ApJL, Bhandari22, Anna-thomas23, Bhandari23, Kumar_2023_MNRAS, Law24, Connor25, Scott25, Dial26, Moroianu26, Muller26, Shin26} where we take median values for repeating sources that exhibit varying time DMs and RMs. 
    \item $\rm DM_{\rm MW,disk}$ is estimated using the Galactic electron density model NE2025 \citep{NE2025}, while $\rm RM_{MW}$, which includes both the Galactic disk and halo contribution, is taken from \cite{Hutschenreuter_2022_A&A}.
    \item $\rm DM_{\rm MW,halo}$ is drawn from a log-normal distribution with mean $\log(30\,\mathrm{pc\,cm}^{-3})$ \citep{Dolag_2015_MNRAS, Yamasaki_2020_ApJ, Cook_2023_ApJ} and standard deviation 0.2\,dex \citep{Yamasaki_2020_ApJ}. 
    \item $\rm DM_{\rm IGM}$ is sampled from the Macquart relation \citep{Macquart_2020_Natur}, accounting for cosmic variance $\sigma_{\rm DM} = Fz^{-1/2}$, with a log-uniform prior on the fluctuation parameter $\log_{10}F \in [-2,0]$ \citep{Baptista_2024_ApJ}. $\rm RM_{\rm IGM}$ is drawn from a Gaussian distribution with mean 0\,rad\,m$^{-2}$ and standard deviation 6\,rad\,m$^{-2}$ \citep{Schnitzeler_2010_MNRAS}.

\end{itemize}

We use the median of the 10,000 realizations for DM$_{\rm host}$ and RM$_{\rm host}$, with lower and upper uncertainties given by the difference between the median and the 16th and 84th percentiles, respectively.

For $\tau_{\rm host,1\,GHz}$, we measure the temporal broadening using {\tt fitburst} \citep{Fonseca_2024_ApJS}, assuming a scattering spectral index of $-4$. The measured scattering timescales are scaled to a reference frequency of 1\,GHz to facilitate comparison with the literature. For the literature sample, we compile scattering measurements  \citep{Price_2019_MNRAS,Lanman_2022_ApJ,Ocker22,Caleb_2023_MNRAS, RBFLOAT,Curtin_2025_ApJ, Scott25,Shin26} and, where necessary, scale them to 1\,GHz using the scattering spectral index reported in the original work. The Milky Way ISM scattering contribution is estimated using the NE2025 model \citep{NE2025} and subtracted from the measurements. The only exception is FRB~20250507A where we constrain an upper limit on the scattering timescale that is smaller than the NE2025 prediction in this line of sight. The resulting host scattering constraint is then rest frame corrected using $\tau_{\rm host,1\,GHz} = \left(\tau_{\rm obs,1\,GHz} - \tau_{\rm MW,1\,GHz}\right)(1+z)^3$.






\bibliography{references}{}
\bibliographystyle{aasjournalv7}



\end{CJK*}
\end{document}

%% file: aff.tex
\newcommand{\NU}{\affiliation{Department of Physics and Astronomy, Northwestern University, Evanston, IL 60208, USA}}

\newcommand{\CIERA}{\affiliation{Center for Interdisciplinary Exploration and Research in Astronomy (CIERA), Northwestern University, 1800 Sherman Avenue, Evanston, IL 60201, USA }}

\newcommand{\MU}{\affiliation{Department of Physics, McGill University, 3600 rue University, Montr\'eal, QC H3A 2T8, Canada}}

\newcommand{\TSI}{\affiliation{Trottier Space Institute, McGill University, 3550 rue University, Montr\'eal, QC H3A 2A7, Canada}}

\newcommand{\UVA}{\affiliation{Anton Pannekoek Institute for Astronomy, University of Amsterdam, Science Park 904, 1098 XH Amsterdam, The Netherlands}}

\newcommand{\DI}{\affiliation{Dunlap Institute for Astronomy and Astrophysics, 50 St. George Street, University of Toronto, ON M5S 3H4, Canada}}

\newcommand{\DAA}{\affiliation{David A. Dunlap Department of Astronomy and Astrophysics, 50 St. George Street, University of Toronto, ON M5S 3H4, Canada}}

\newcommand{\UCSC}{\affiliation{Department of Astronomy and Astrophysics, University of California, Santa Cruz, 1156 High Street, Santa Cruz, CA 95060, USA}}

\newcommand{\SKAO}{\affiliation{SKA Observatory, 26 Dick Perry Ave, Kensington WA 6151 Australia}}

\newcommand{\WVUPHAS}
{\affiliation{Department of Physics and Astronomy, West Virginia University, PO Box 6315, Morgantown, WV 26506, USA }}

\newcommand{\WVUGWAC}
{\affiliation{Center for Gravitational Waves and Cosmology, West Virginia University, Chestnut Ridge Research Building, Morgantown, WV 26505, USA}}

\newcommand{\MITK}
{\affiliation{MIT Kavli Institute for Astrophysics and Space Research, Massachusetts Institute of Technology, 77 Massachusetts Ave, Cambridge, MA 02139, USA}}

\newcommand{\MITP}
{\affiliation{Department of Physics, Massachusetts Institute of Technology, 77 Massachusetts Ave, Cambridge, MA 02139, USA}}

\newcommand{\LAM}
{\affiliation{Laboratoire d'Astrophysique de Marseille, Aix-Marseille Univ., CNRS, CNES, Marseille, France}}

\newcommand{\PI}
{\affiliation{Perimeter Institute of Theoretical Physics, 31 Caroline Street North, Waterloo, ON N2L 2Y5, Canada}}

\newcommand{\YORK}
{\affiliation{Department of Physics and Astronomy, York University, 4700 Keele Street, Toronto, ON MJ3 1P3, Canada}}

\newcommand{\MIBR}
{\affiliation{Miller Institute for Basic Research, University of California, Berkeley, CA 94720, United States}}

\newcommand{\UCBASTRO}
{\affiliation{Department of Astronomy, University of California, Berkeley, CA 94720, United States}}

\newcommand{\PRINCETON}
{\affiliation{Department of Astrophysical Sciences, Princeton University, 4 Ivy Lane, Princeton, NJ 08544, USA}}

\newcommand{\IPMU}{\affiliation{Kavli Institute for the Physics and Mathematics of the Universe (Kavli IPMU), 5-1-5 Kashiwanoha, Kashiwa, 277-8583, Japan}}

\newcommand{\NAOJ}{\affiliation{Division of Science, National Astronomical Observatory of Japan,2-21-1 Osawa, Mitaka, Tokyo 181-8588, Japan}}

\newcommand{\ASTRON}{\affiliation{ASTRON, Netherlands Institute for Radio Astronomy, Oude Hoogeveensedijk 4, 7991 PD Dwingeloo, The Netherlands}}

\newcommand{\UCh}{\affiliation{Department of Astronomy and Astrophysics, University of Chicago, William Eckhart Research Center, 5640 South Ellis Avenue, Chicago, IL 60637, USA}}

\newcommand{\CCAPS}{\affiliation{Department of Astronomy and Cornell Center for Astrophysics and Planetary Science, Cornell University, Ithaca, NY 14853, USA}}

\newcommand{\CALTECH}{\affiliation{Division of Physics, Mathematics, and Astronomy, California Institute of Technology, Pasadena, CA 91125, USA}}

\newcommand{\CFA}{\affiliation{Center for Astrophysics | Harvard \& Smithsonian, 60 Garden St, Cambridge, MA 02138, USA}}

%% file: authors.tex
\author[0000-0002-9363-8606]{Y.~Dong (董雨欣)}
\CIERA
\NU
\email[show]{yuxin.dong@northwestern.edu}

\author[0000-0002-7374-935X]{W.~Fong}
\CIERA
\NU
\email{wfong@northwestern.edu}

\author[0000-0003-0510-0740]{K.~Nimmo}
\altaffiliation{NASA Einstein Fellow}
\CIERA
\CFA
\email{knimmo@northwestern.edu} 

\author[0000-0001-7599-6664]{N.~Loudas}
\PRINCETON
\email{loudas@princeton.edu}

\author[0000-0001-5908-3152]{B.~C.~Andersen}
\UCSC
\email{banders8@ucsc.edu}

\author[0000-0002-3980-815X]{S.~Andrew}
\MITK
\MITP
\email{shiona@mit.edu}

\author[0009-0001-0983-623X]{A.~Cai}
\CIERA
\NU
\email{acai@u.northwestern.edu}

\author[0000-0002-2878-1502]{S.~Chatterjee}
\CCAPS
\email{shami.chatterjee@cornell.edu}

\author[0000-0001-6422-8125]{A.M.~Cook}
\MU
\TSI
\UVA
\email{amanda.cook@mail.mcgill.ca}

\author[0000-0002-8376-1563]{A.P.~Curtin}
\MU
\TSI
\UVA
\email{alice.curtin@mcgill.ca}

\author[0000-0003-0307-9984]{T.~Eftekhari}
\CIERA
\email{teftekhari@northwestern.edu}

\author[0000-0002-3382-9558]{B.M.~Gaensler}
\UCSC
\DI
\DAA
\email{gaensler@ucsc.edu}

\author[0000-0003-2317-1446]{J.~Hessels}
\MU
\TSI
\UVA
\ASTRON
\email{jason.hessels@mcgill.ca}

\author[0000-0001-9345-0307]{V.M.~Kaspi}
\MU
\TSI
\email{victoria.kaspi@mcgill.ca}

\author[0009-0004-4176-0062]{A.~Khan}
\MU
\TSI
\email{afrasiyab.khan@mcgill.ca}

\author[0000-0002-4209-7408]{C.~Leung}
\MIBR
\UCBASTRO
\email{calvin_leung@berkeley.edu}

\author[0000-0003-4584-8841]{L.~Mas-Ribas}
\UCSC
\email{lmr@ucsc.edu}

\author[0000-0002-4279-6946]{K.~Masui}
\MITK
\MITP
\email{kmasui@mit.edu}

\author[0000-0003-1936-9062]{A.~Moroianu}
\UVA
\ASTRON
\email{a.m.moroianu@uva.nl}

\author[0000-0002-8912-0732]{A.B.~Pearlman}
\altaffiliation{NASA Hubble Fellow}
\email{aaron.b.pearlman@mit.edu}
\MITK
\MU
\TSI

\author[0000-0002-7738-6875]{J.X.~Prochaska}
\UCSC
\IPMU
\NAOJ
\email{xavier@ucolick.org}

\author[0000-0002-4623-5329]{M.W.~Sammons}
\LAM
\email{mawson.sammons@lam.fr}

\author[0000-0002-7374-7119]{P.~Scholz}
\email{pscholz@yorku.ca}
\YORK

\author[0000-0002-6823-2073]{K.~Shin}
\CALTECH
\email{kaitshin@caltech.edu}

\author[0000-0003-3801-1496]{S.~Simha}
\CIERA
\UCh
\email{sunil.simha@northwestern.edu}

\author[0009-0001-3334-9482]{M.~Woodland}
\UCSC
\email{miwoodla@ucsc.edu}

\author[0000-0002-1945-2299]{Z.~Zhuang}
\email{zhuyun.zhuang@northwestern.edu} 
\CIERA

%% file: frb_basics.tex
\begin{deluxetable*}{l|cccccccccccc}[t!]
\tablewidth{\textwidth}
\tablecaption{Basic FRB burst and host information of all eight dwarf candidates in the sample. \label{tab:basics}}
\tablecolumns{8}
\tablewidth{0pt}
\tablehead{
\colhead{FRB} &
\colhead{R.A.} & 
\colhead{Dec.} &
\colhead{$a_{\rm err}$} &
\colhead{$b_{\rm err}$} &
\colhead{$\theta$} &
\colhead{$z$} &
\colhead{Host R.A.} &
\colhead{Host Dec.} &
\colhead{$m_r$} &
\colhead{$r$-band Luminosity} & 
\colhead{$P(O|x)$} \\
\colhead{} &
\colhead{(J2000)} &
\colhead{(J2000)} &
\colhead{arcsec} &
\colhead{arcsec} & 
\colhead{deg} &
\colhead{} &
\colhead{(J2000)} &
\colhead{(J2000)} &
\colhead{mag} & 
\colhead{log($L/L_\odot$)} &
\colhead{} 
}
\startdata
\textbf{20220529A}$^{a}$ & \ra{01}{16}{25} & \dec{20}{37}{57} & 1.39 & 0.64 & -12.6 & 0.1839 & \ra{01}{16}{25} & \dec{+20}{37}{58} & 21.64 & 9.21 & 0.99 \\
\textbf{20250206A} & \ra{22}{34}{41} & \dec{+12}{10}{45} & 1.88 & 0.76 & 101.54 & 0.1180 & \ra{22}{34}{41} & \dec{+12}{10}{45} & 21.70 & 8.76 & 0.99 \\ 
\textbf{20250227A} & \ra{07}{33}{31} & \dec{+15}{13}{24} & 0.45 & 0.12 & 12.72 & 0.0314 & \ra{07}{33}{30} & \dec{+15}{13}{35} & 16.97 & 9.45 & 0.99 \\
\textbf{20250316A}$^{b}$ & \ra{12}{09}{44} & \dec{+58}{50}{57} & 0.07 & 0.06 & -0.26 & 0.0065 & \ra{12}{09}{47} & \dec{+58}{50}{57} & 14.60 & 9.02 & 1.00 \\
20250410E & \ra{00}{56}{10} & \dec{-05}{08}{19} & 1.8 & 0.9 & 16.64 & 0.4980 & \ra{00}{56}{10} & \dec{-05}{08}{19} & 23.56 & 9.43 & 0.93 \\
20250507A & \ra{23}{58}{32} & \dec{+42}{56}{32} & 0.3 & 0.04 & 0.60 & 0.1204 & \ra{23}{58}{32} & \dec{+42}{56}{32} & 20.29 & 9.34 & 0.99 \\
20250515A & \ra{02}{51}{29} & \dec{+08}{02}{10} & 0.4 & 0.1 & 14.54 & 0.0625 & \ra{02}{51}{30} & \dec{+08}{02}{13} & 18.75 & 9.36 & 0.98 \\
\textbf{20250704A} & \ra{07}{13}{48} & \dec{+24}{57}{36} & 0.26 & 0.07 & 11.60 & 0.1130 & \ra{07}{13}{48} & \dec{+24}{57}{37} & 22.82 & 8.27 & 0.90 \\
\enddata
\tablecomments{This represents all hosts which met our luminosity threshold to be considered a dwarf candidate. Semi-major and semi-minor axes and the position angles of the $1\sigma$ FRB localization ellipses are provided as $a_{\rm err}$ and $b_{\rm err}$ in arcseconds, and $\theta$ in degrees, measured east of north. The redshifts are derived from spectroscopic observations in this work unless otherwise specified. We bold the FRB names of the confirmed dwarf host galaxies identified in this work; the remaining entries are new hosts that were rejected as dwarf candidates (See Section~\ref{sec:dwarfidentification}). \\
$^{a}$ Repeater. Redshift and host position from \cite{Li26}. FRB position and localization from \cite{Pandhi26}. \\
$^{b}$ FRB position and localization from \cite{RBFLOAT}.}
\end{deluxetable*}


%% file: hostproperties.tex
\begin{deluxetable*}{l|cccccccc}
\tabletypesize{\footnotesize}
\setlength{\tabcolsep}{8pt}
\tablecolumns{10}
\renewcommand{\arraystretch}{1.25}
\tablewidth{\textwidth}
\tablecaption{Stellar Population Properties from \texttt{Prospector}\label{tab:host_prop}}
\tablehead{
\colhead{FRB} &
\colhead{log(M$_{\rm F}$/M$_{\odot}$)} &
\colhead{log($M_*/M_\odot$)} &
\colhead{${\rm SFR}_{\rm 0-100 Myr}$} &
\colhead{log($Z_*/Z_\odot$)} &
\colhead{log($Z_\textrm{gas}/Z_\odot$)} &
\colhead{A$_{\rm V, young}$} &
\colhead{A$_{\rm V, old}$}  &
\colhead{$t_{\rm m}$} \\
\colhead{} &
\colhead{} &
\colhead{} &
\colhead{[$M_\odot$~yr$^{-1}$]} & 
\colhead{} &
\colhead{} &
\colhead{[mag]} &
\colhead{[mag]} & 
\colhead{[Gyr]}
}
\startdata
20220529A & $9.24^{+0.13}_{-0.13}$ & $9.12^{+0.12}_{-0.11}$ & $0.87^{+0.61}_{-0.40}$ & $-1.69^{+0.15}_{-0.15}$ & $-0.49^{+0.08}_{-0.08}$ & $0.35^{+0.28}_{-0.21}$ & $2.76^{+0.46}_{-0.31}$ & $2.02^{+1.45}_{-1.40}$ \\
20250206A & $8.83^{+0.06}_{-0.07}$ & $8.61^{+0.05}_{-0.06}$ & $0.03^{+0.02}_{-0.01}$ & $-1.91^{+0.20}_{-0.06}$ & $-0.38^{+0.16}_{-0.18}$ & $0.08^{+0.11}_{-0.06}$ & $0.08^{+0.10}_{-0.06}$ & $5.86^{+0.63}_{-0.84}$   \\
20250227A & $9.12^{+0.08}_{-0.08}$ & $8.92^{+0.07}_{-0.07}$ & $0.51^{+0.14}_{-0.14}$ & $-0.75^{+0.09}_{-0.08}$ & $-0.18^{+0.22}_{-0.32}$ & $0.09^{+0.08}_{-0.06}$ & $0.10^{+0.08}_{-0.06}$ & $5.24^{+0.73}_{-0.92}$ \\
20250316A & $9.33^{+0.03}_{-0.03}$ & $9.11^{+0.03}_{-0.03}$ & $0.04^{+0.02}_{-0.02}$ & $-0.92^{+0.05}_{-0.05}$ & $-0.44^{+0.16}_{-0.14}$ & $0.01^{+0.01}_{-0.01}$ & $0.01^{+0.01}_{-0.01}$ & $6.95^{+0.57}_{-0.65}$ \\ 
20250704A & $8.31^{+0.08}_{-0.11}$ & $8.09^{+0.08}_{-0.11}$ & $0.02^{+0.02}_{-0.01}$ & $-1.68^{+0.33}_{-0.21}$ & $-1.01^{+0.72}_{-0.57}$ & $0.11^{+0.18}_{-0.08}$ & $0.12^{+0.16}_{-0.08}$ & $5.97^{+0.55}_{-0.85}$ \\
\hline
\multicolumn{9}{c}{Rejected Dwarf Candidates} \\
\hline
20250410E & $9.70^{+0.15}_{-0.13}$ & $9.48^{+0.15}_{-0.12}$ & $0.26^{+0.18}_{-0.11}$ & $-1.42^{+0.18}_{-0.17}$ & $-0.76^{+0.15}_{-0.13}$ & $0.27^{+0.26}_{-0.17}$ & $0.29^{+0.27}_{-0.19}$ & $4.75^{+0.73}_{-0.36}$ \\
20250507A & $9.50^{+0.07}_{-0.07}$ & $9.28^{+0.06}_{-0.07}$ & $0.17^{+0.16}_{-0.08}$ & $-0.93^{+0.24}_{-0.12}$ & $-0.32^{+0.10}_{-0.11}$ & $0.33^{+0.26}_{-0.15}$ & $0.33^{+0.19}_{-0.13}$ & $6.74^{+0.72}_{-0.85}$ \\
20250515A & $9.74^{+0.05}_{-0.04}$ & $9.53^{+0.04}_{-0.03}$ & $0.42^{+0.16}_{-0.09}$ & $-0.15^{+0.04}_{-0.17}$ & $-1.51^{+0.15}_{-0.10}$ & $0.01^{+0.01}_{-0.01}$ & $0.01^{+0.01}_{-0.01}$ & $7.95^{+1.65}_{-0.70}$ \\
\enddata
\tablecomments{Median and 68$\%$ confidence intervals of the dwarf host stellar population properties. log(M$_{\rm F}$/M$_{\odot}$) is the total mass formed and log(M$_{*}$/M$_{\odot}$) is stellar mass formed. ${\rm SFR}_{\rm 0-100 Myr}$ is star formation within the last 100~Myr, log(Z$_*$/Z$_{\odot}$) is the stellar metallicity, and log($Z_\textrm{gas}/Z_\odot$) is the gas-phase metallicity. A$_{\rm V, young}$ and A$_{\rm V, old}$ are the magnitudes of dust extinction for young and old stars, respectively. $t_{\rm m}$ is the mass-weighted age.}
\end{deluxetable*}

%% file: emlines.tex
\begin{deluxetable*}{l|ccccc}
\tablecolumns{6}
\tablewidth{\textwidth}
\tabletypesize{\normalsize}
\caption{Nebular Emission-line Fluxes for Low-mass FRB Hosts}
\tablehead{
\colhead{FRB} & 
\colhead{H$\alpha$} &
\colhead{H$\beta$} &
\colhead{[OIII]} & 
\colhead{[NII]} & 
\colhead{$12+\log(\mathrm{O/H})$} \\
\colhead{} & 
\colhead{} & 
\colhead{} & 
\colhead{5007 \AA} & 
\colhead{6584 \AA} &
\colhead{}
}  
\startdata
20220529A & 1.645$^{+0.068}_{-0.069}$ & 0.288$^{+0.015}_{-0.015}$ & 0.793$^{+0.030}_{-0.031}$ & 0.370$^{+0.015}_{-0.015}$ & 8.504$^{+0.004}_{-0.003}$ \\ [3pt] 
20250206A & 1.219$^{+0.059}_{-0.054}$ & 0.338$^{+0.018}_{-0.016}$ & 0.298$^{+0.024}_{-0.030}$ & 0.036$^{+0.003}_{-0.002}$ & 8.400$^{+0.012}_{-0.012}$ \\ [3pt] 
20250227A & 36.448$^{+1.096}_{-1.105}$ & 11.991$^{+0.580}_{-0.524}$ & 11.927$^{+0.462}_{-0.459}$ & 7.346 $^{+0.250}_{-0.297}$ & 8.600$^{+0.002}_{-0.003}$ \\ [3pt] 
20250316A & 563.248$^{+13.990}_{-14.854}$ & 158.698$^{+5.329}_{-5.655}$ & 128.129$^{+3.548}_{-3.620}$ & 137.405$^{+3.819}_{-3.685}$ & 8.639$^{+0.002}_{-0.002}$ \\ [3pt]
20250704A & 0.228$^{+0.034}_{-0.035}$ & $<0.3$ & 0.200$^{+0.156}_{-0.079}$ & $<0.09$ & $<$8.699 \\ [3pt] 
\hline
\multicolumn{6}{c}{Rejected Dwarf Candidates} \\
\hline
20250410E & 0.235$^{+0.053}_{-0.036}$ & 0.066$^{+0.009}_{-0.008}$ & 0.057$^{+0.017}_{-0.017}$ & $<0.061$ & $<$8.640 \\ [3pt]
20250507A & 11.491$^{+0.295}_{-0.309}$ & 3.689$^{+0.122}_{-0.120}$ & 7.416$^{+0.226}_{-0.231}$ & 1.901$^{+0.048}_{-0.052}$ & 8.505$^{+0.001}_{-0.001}$ \\ [3pt]
20250515A & 11.205$^{+0.335}_{-0.348}$ & 3.449$^{+0.135}_{-0.143}$ & 4.088$^{+0.119}_{-0.138}$ & 3.036$^{+0.096}_{-0.099}$ & 8.612$^{+0.002}_{-0.002}$ \\
[3pt]
\enddata
\tablecomments{Fluxes are fitted using \texttt{Prospector} in units of $10^{-16} \ \rm erg \ s^{-1} \ cm^{-2}$ and corrected for Galactic extinction. The H$\beta$ and [NII] flux limits for FRBs\,20250410E and 20250704A correspond to 3$\sigma$. The uncertainties are from MCMC sampling of the line flux measurements only and do not include systematic uncertainties in the metallicity calibration. \\}
\label{tab:lines}
\end{deluxetable*}

%% file: frb_radio.tex
\begin{deluxetable*}{l|ccccccc}[t!]
\linespread{0.9}
\tabletypesize{\normalsize}
\tablecaption{Measured and Inferred Radio Properties of the Eight FRB Host Galaxies in This Work \label{tab:frb_radio}}
\tablecolumns{8}
\tablewidth{0pt}
\tablehead{
\colhead{FRB} &
\colhead{DM$_\mathrm{obs}$} &
\colhead{RM$_\mathrm{obs}$} &
\colhead{$\tau_{600\,\mathrm{MHz,obs}}$} &
\colhead{DM$_\mathrm{host, rf}$} &
\colhead{RM$_\mathrm{host, rf}$} &
\colhead{$\tau_\mathrm{host, rf, 1GHz}$}  \\
\colhead{} &
\colhead{(pc~cm$^{-3}$)} &
\colhead{(rad~m$^{-2}$)} & 
\colhead{(ms)} &
\colhead{(pc~cm$^{-3}$)} &
\colhead{(rad~m$^{-2}$)} & 
\colhead{(ms)} 
}
\startdata
\textbf{20220529A}$^{a}$ & $245.6\pm0.8$ & $-0.4\pm22.6$ & $0.93\pm0.77$ &  $62_{-62}^{+36}$  & $74 \pm 13 $   & $0.20\pm0.17$  \\
\textbf{20250206A} & 207.11708(8) & $-63.13$(9) & 0.0541(5) & $61_{-57}^{+28}$  & $-31\pm15$ & $0.0096$(1) \\ 
\textbf{20250227A} & 156.6007(2) & $+5.08$(7) & 0.074(1) &  $26_{-26}^{+18}$  & $-32 \pm 14$  & $0.0082$(1) \\
\textbf{20250316A}$^{b}$ & 161.82(2) & +16.79(85) & 0.405(1) &  $95_{-19}^{+13}$  & $-1 \pm 7$   & 0.0534(1)  \\
20250410E$^{c}$ & 619.082(2) & - & 2.86(4) & $292_{-225}^{+94}$ & -  &  $1.25(2)$ \\
20250507A & 212.6440(2) & $-0.094$(5) & $<0.002$ &  $34_{-34}^{+28}$  & $-56\pm15$  &  $<0.0004$ \\
20250515A & 169.3728(6) & $-1.55$(9) & $<0.1$ & $64_{-35}^{+20}$ & $4 \pm 9$ & $<0.02$  \\
\textbf{20250704A} & 356.4452(5) & $-96.70(3)$ & 0.664(2) &  $185_{-56}^{+29}$ & $-101 \pm 19$  & $0.1136(4)$ \\
\enddata
\tablecomments{The confirmed dwarfs are shown in bold, while the remaining FRB hosts are those rejected as dwarf galaxies. We denote rest-frame parameters with the subscript ``rf''. The scattering timescale from the host, $\tau_{\rm host}$, is given at 1~GHz in the rest frame. \\
$^{a}$ Median of the measured values, where we consider the absolute values of the RM and we ignore upper limits of scattering in the median determination, so it is likely higher than the truth in \citet{Pandhi26}. The uncertainties reflect the standard deviation of observed values in \citet{Pandhi26}. Note that we exclude the RM flare presented in \citet{Li26}.\\
$^{b}$ Measured values taken from \citet{RBFLOAT}. \\
$^{c}$ Calibration issues prevent us from measuring a convincing RM.}
\end{deluxetable*}

%% file: correlation_tests.tex
\begin{deluxetable*}{lccc|ccc|ccc}
\tablecaption{Kendall's $\tau$ correlation statistics and two-sided
permutation-test $p$-values.\label{tab:tau_pvalues}}
\tablehead{
\colhead{Selection} &
\multicolumn{3}{c|}{$\mathrm{DM}_{\mathrm{host}}$} &
\multicolumn{3}{c|}{$\mathrm{RM}_{\mathrm{host}}$} &
\multicolumn{3}{c}{$\tau_{\mathrm{host,1\,GHz}}$}}
\startdata
& \multicolumn{1}{c}{$N$}
& \multicolumn{1}{c}{$\tau_{\rm k}$}
& \multicolumn{1}{c|}{$p$}
& \multicolumn{1}{c}{$N$}
& \multicolumn{1}{c}{$\tau_{\rm k}$}
& \multicolumn{1}{c|}{$p$}
& \multicolumn{1}{c}{$N$}
& \multicolumn{1}{c}{$\tau_{\rm k}$}
& \multicolumn{1}{c}{$p$} \\
All
& 54 & $-0.01^{+0.04}_{-0.05}$ & 0.941
& 37 & $-0.15^{+0.03}_{-0.03}$ & 0.226
& 24 & $0.31^{+0.04}_{-0.03}$ & 0.052 \\
No PRS
& 50 & $0.11^{+0.05}_{-0.05}$ & 0.281
& 33 & $0.03^{+0.04}_{-0.04}$ & 0.810
& 20 & $0.45^{+0.05}_{-0.05}$ & 0.008 \\
\enddata

\tablecomments{
$N$ is the number of objects included in each correlation test.
The $\tau_{\rm k}$ values are the median Kendall-equivalent statistics from 1000 Monte Carlo realizations, with uncertainties corresponding to the 16th and 84th percentiles. The $p$-values are two-sided permutation-test probabilities for the null hypothesis of no correlation with global host metallicity, with the censoring structure preserved. We find a marginal correlation between $\tau_\mathrm{host,1,GHz}$ and metallicity, with a significance of $2\sigma$ for a $p$-value threshold of 0.01. After trial correction, this corresponds to $1\sigma$. Excluding FRB hosts with known PRS, the correlation increases to $2.6\sigma$ for the same threshold, or $2\sigma$ after trial correction. } 
\end{deluxetable*}

%% file: photometry.tex
\startlongtable
\begin{deluxetable*}{l|ccccc}
\tabletypesize{\footnotesize}
\tablecolumns{7}
\tablewidth{0pc}
\tablecaption{Dwarf FRB Host and Candidate Galaxy Photometry}
\label{tab:phot}
\tablehead{
\colhead{FRB} &
\colhead{Host R.A.} &
\colhead{Host Dec.} &
\colhead{Instrument or Survey} &
\colhead{Filter} &
\colhead{AB Mag} \\
\colhead{} &
\colhead{(J2000)} & 
\colhead{(J2000)} 
}
\startdata
20220529A & \ra{01}{16}{25.08} & \dec{+20}{37}{57.58} & Legacy & $g$ & 21.89 $\pm$ 0.04 \\
 &  &  & Legacy  & $r$ & 21.45 $\pm$ 0.04 \\
 &  &  & Legacy & $z$ & 21.09 $\pm$ 0.06 \\
20250206A & \ra{22}{34}{41.14} & \dec{+12}{10}{45.29} & Legacy & $g$ & 22.01 $\pm$ 0.04 \\
 &  &  & Legacy  & $r$ & 21.61 $\pm$ 0.08 \\
 &  &  & Legacy & $z$ & 21.58 $\pm$ 0.11 \\
20250227A & \ra{07}{33}{30.17} & \dec{+15}{13}{34.70} & Legacy & $g$ & 17.60 $\pm$ 0.004 \\
 &  &  & Legacy  & $r$ & 17.27 $\pm$ 0.005 \\
 &  &  & Legacy & $z$ & 17.35 $\pm$ 0.009 \\
 &  &  & WISE & $W1$ & 17.88 $\pm$ 0.07\\
&  &  & WISE & $W2$ & 18.53 $\pm$ 0.19 \\
20250316A & \ra{12}{09}{47.32} & \dec{+58}{50}{57.10} & Legacy & $g$ & 14.15 $\pm$ 0.0002 \\
 &  &  & Legacy  & $r$ & 13.86 $\pm$ 0.0004 \\
 &  &  & Legacy & $z$ & 14.05 $\pm$ 0.0015 \\
 &  &  & WISE & $W1$ & 14.62 $\pm$ 0.007 \\
&  &  & WISE & $W2$ & 15.24 $\pm$ 0.016 \\
&  &  & WISE & $W3$ & 14.14 $\pm$ 0.027 \\
&  &  & WISE & $W4$ & 12.92 $\pm$ 0.111 \\
20250410E & \ra{00}{56}{10.30} & \dec{-05}{08}{18.77} & Gemini/GMOS-S & $g$ & 23.45 $\pm$ 0.06 \\
20250507A & \ra{23}{58}{31.49} & \dec{+42}{56}{32.43} & Pan-STARRS & $g$ & 20.95 $\pm$ 0.04 \\
 &  &  & Pan-STARRS & $r$ & 20.44 $\pm$ 0.03 \\
 &  &  & Pan-STARRS & $i$ & 20.15 $\pm$ 0.02 \\
 &  &  & Pan-STARRS & $z$ & 20.04 $\pm$ 0.04 \\
&  &  & Pan-STARRS & $y$ & 19.89 $\pm$ 0.08 \\
20250515A & \ra{02}{51}{29.66} & \dec{+08}{02}{12.99} & Legacy & $g$ & 18.66 $\pm$ 0.01 \\
 &  &  & Legacy & $r$ & 18.27 $\pm$ 0.01 \\
 &  &  & Legacy & $z$ & 18.03 $\pm$ 0.02 \\
 &  &  & WISE & $z$ & 18.65 $\pm$ 0.10 \\
&  &  & WISE & $y$ & 18.99 $\pm$ 0.32 \\
20250704A & \ra{07}{13}{47.91} & \dec{+24}{57}{36.61} & Gemini/GMOS-S & $r$ & 22.65 $\pm$ 0.10 \\
 &  &  & MMT/MMIRS & $J$ & 22.88 $\pm$ 0.21 \\
 \hline
 \enddata
\tablecomments{The photometry points are collected from archival surveys and from this work. Magnitudes are corrected for Galactic extinction.}
\end{deluxetable*}